%% file: proto.tex
\documentclass[fleqn,usenatbib,useAMS]{mnras}

\usepackage{graphicx}   
\usepackage{amsmath}    
\usepackage{amssymb}    
\usepackage{multicol}   
\usepackage{bm}            
\usepackage{float}
\usepackage{comment}
\usepackage{tabularx}
\usepackage{hyperref}
\usepackage{threeparttable}
\newcommand{\width}{.43\textwidth}
\newcommand{\twidth}{.47\textwidth}
\newcommand{\centered}[1]{\begin{tabular}{l} #1 \end{tabular}}
\newcommand{\Tstrut}{\rule{0pt}{2.6ex}}
\newcommand{\Bstrut}{\rule[-0.9ex]{0pt}{0pt}}

\usepackage[T1]{fontenc}
\usepackage{ae,aecompl}
\usepackage{newtxtext,newtxmath}
\usepackage{cite}
\title[Overdensities around protocluster cores]{Overdensities of Submillimetre-Bright Sources around Candidate Protocluster Cores Selected from the South Pole Telescope Survey}

\author[Wang et al.]
{George Wang,$^{1}$
Ryley Hill,$^{1}$
S.\ C.\ Chapman,$^{1,2,3}$
A.~Wei\ss,$^{4}$
Douglas Scott,$^{1}$\newauthor
Manuel Aravena,$^{5}$
Melanie Ann Archipley,$^{6}$
Matthieu B{\'e}thermin,$^{7}$
Carlos De Breuck,$^{8}$\newauthor
R.E.A. Canning,$^{9}$
Chenxing Dong,$^{10}$
W. B. Everett,$^{11}$
Anthony Gonzalez,$^{10}$\newauthor
Thomas R. Greve,$^{12,13,14}$
Christopher C.~Hayward,$^{15}$
Yashar Hezaveh,$^{15, 16}$
D. P. Marrone,$^{17}$\newauthor
Sreevani Jarugula,$^{6}$
Kedar A. Phadke,$^{6}$
Cassie A. Reuter,$^{6}$
Justin S. Spilker,$^{18, \dagger}$\newauthor
Joaquin D. Vieira$^{6}$\\
$^{1}$Department of Physics and Astronomy, University of British Columbia,
6225 Agricultural Road, Vancouver, V6T 1Z1, Canada\\
$^{2}$National Research Council, Herzberg Astronomy and Astrophysics, 5071
West Saanich Road, Victoria, V9E 2E7, Canada\\
$^{3}$Department of Physics and Atmospheric Science, Dalhousie University,
B3H 4R2, Halifax, Canada\\
$^{4}$Max-Planck-Institut f{\"u}r Radioastronomie, Auf dem  H{\"u}gel 69, D-53121, Bonn, Germany\\
$^{5}$N\'ucleo de Astronom\'{\i}a, Facultad de Ingenier\'{\i}a y Ciencias, Universidad Diego Portales, Av. Ej{\'e}rcito 441, Santiago, Chile\\
$^{6}$Department of Astronomy, University of Illinois, 1002 West Green St., Urbana, IL 61801, USA\\
$^{7}$Laboratoire d'Astrophysique de Marseille, 38 rue Fr{\'e}d{\'e}ric Joliot-Curie, Marseille, 13013, France\\
$^{8}$European Southern Observatory, Karl Schwarzschild Stra\ss e 2, Garching, D-85748, Germany\\
$^{9}$Stanford University, 382 Via Pueblo Mall Stanford, CA 94305-4013\\
$^{10}$Department of Astronomy, University of Florida, Gainesville, FL 32611, USA\\
$^{11}$Center for Astrophysics and Space Astronomy and Department of Astrophysical and Planetary Sciences, University of Colorado, Boulder, CO, USA 80309\\
$^{12}$Cosmic Dawn Center, Holbergsgade 14, Copenhagen, DK-1057, Denmark\\
$^{13}$Department of Physics and Astronomy, University College London, Gower Street, London WC1E 6B, UK\\
$^{14}$DTU-Space, National Space Institute, Technical University of
Denmark, Elektrovej 327, DK-2800 Kgs. Lyngby, Denmark\\
$^{15}$Center for Computational Astrophysics, Flatiron Institute, 162
Fifth Avenue, New York, NY 10010, USA\\
$^{16}$D{\'e}partement de Physique, Universit{\'e} de Montr{\'e}al,
Montreal, Quebec, Canada H3T 1J4\\
$^{17}$Steward Observatory, University of Arizona, 933 North Cherry Avenue, Tucson, AZ 85721, USA\\
$^{18}$Department of Astronomy, University of Texas at Austin, 2515 Speedway, Stop C1400, Austin, TX 78712, USA\\
$^{\dagger}$NHFP Hubble Fellow
}

\pubyear{2020}
\begin{document}
\label{firstpage}
\begin{abstract}
\pagerange{\pageref{firstpage}--\pageref{lastpage}}
\maketitle
\vspace{-0.7cm}
We present APEX-LABOCA 870-$\mu$m observations of the fields surrounding the nine brightest, high-redshift, unlensed objects discovered in the South Pole Telescope's (SPT) 2500\,deg$^{2}$ survey. Initially seen as point sources by SPT's 1-arcmin beam, the 19-arcsec resolution of our new data enables us to deblend these objects and search for submillimetre (submm) sources in the surrounding fields. We find a total of 98 sources above a threshold of 3.7$\sigma$ in the observed area of 1300\,arcmin$^2$, where the bright central cores resolve into multiple components. After applying a radial cut to our LABOCA sources to achieve uniform sensitivity and angular size across each of the nine fields, we compute the cumulative and differential number counts and compare them to estimates of the background, finding a significant overdensity of $\delta\,{\approx}\,$10 at $S_{870}\,{=}\,14$\,mJy. The large overdensities of bright submm sources surrounding these fields suggest that they could be candidate protoclusters undergoing massive star-formation events. Photometric and spectroscopic redshifts of the unlensed central objects range from $z\,{=}\,$3 to 7, implying a volume density of star-forming protoclusters of approximately 0.1\,Gpc$^{-3}$. If the surrounding submm sources in these fields are at the same redshifts as the central objects, then the total star-formation rates of these candidate protoclusters reach 10,000\,M$_{\odot}$\,yr$^{-1}$, making them much more active at these redshifts than what has been seen so far in both simulations and observations.
\end{abstract}


\section{Introduction}

Massive galaxy clusters are now identified as early as $z$\,$\approx$\,2 by searching for overdensities of red, early-type galaxies \citep[e.g.][]{Gladders2000, stanford2005, Wilson2006, Eisenhardt2008, papovich2010, tanaka2010, gobat2011, wylezalek2014, noirot2016}. These methods require near- and mid-infrared (IR) observations for redshifts $z$\,${\gtrsim}$\,1, and have been successful at identifying structures in the early Universe. Other well-established observational signatures, such as X-rays emitted by hot intra-cluster gas \citep[e.g.,][]{bohringer2000,Fassbender2011,pacaud2016} and the Sunyaev-Zeldovich effect \citep[e.g.,][]{Hasselfield2013,planck2013-p15,bleem2015,huang2019} have confirmed these distant structures and pioneered surveys for cluster identification. Cosmological simulations suggest a large variation in halo growth histories; overdensities at much higher redshifts correspond to both today's largest galaxy clusters and individual massive galaxies \citep[e.g.,][]{springel2005,overzier2009}. These protocluster regions are built up hierarchically, and now contain mostly ``red and dead'' galaxies. At some stages in their evolution, they are expected to contain extremely active star-forming galaxies \citep[e.g.,][]{miley2008, miller2015, Chiang2017}, which implies strong emission at submm wavelength, before quenching of the star formation takes place \citep[e.g.,][]{lewis2002,boselli2006}.

Locating such protoclusters is challenging since candidates found in extensive mapping surveys with submm-to-millimetre (mm) wavelength telescopes have relatively modest beam sizes, and can only be confirmed as genuine protoclusters through targeted follow-up observations. Despite these challenges, several protoclusters have been identified at $z\,{\gtrsim}\,2$, each containing up to a dozen high star-formation rate (SFR) galaxies \citep[e.g.][]{chapman2009,tamura2009,dannerbauer2014,casey2015,chiang2015,umehata2015,hung2016, Greenslade2018, Cheng2019, Kneissl2018, lacaille2019}. However, the selection criteria for these objects tend to vary dramatically from system to system, making it challenging to derive abundances and say anything conclusive concerning the density of early-forming clusters.

The SPT-SZ survey with the South Pole Telescope (SPT) has identified numerous, bright (flux density at 1.4-mm, $S_{1.4}>25$\,mJy) sources unresolved by the telescope's 1-arcmin beam \citep{vieira2010, mocanu2013, Everett2020}. With the help of ground- and space-based facilities, such as the Atacama Large Millimeter Array (ALMA), {\it Herschel}, and APEX, a majority of these sources have been confirmed to be strong gravitational lenses, with typical sizes of 2\,arcsec \citep[][]{vieira2013,spilker2016}. However, Chapman et al.\ (in prep) have demonstrated that a subset of $\sim$\,10\,per cent resolves into multiple sources seen in ALMA 3-mm, and 850-$\mu$m observations, finding several sources at the same redshift in each case. The SPT sources are even resolved at the relatively coarse spatial resolution of \textit{Herschel}, and ground-based bolometer cameras like LABOCA. These SPT sources are unlensed and represent collections of high-SFR galaxies packed within a relatively small solid angle. These {\it unlensed} sources are candidate protocluster cores (Chapman et al.\ in prep.), of which the now well-studied $z\,{=}\,4.30$ SPT2349-56 \citep{Miller2018, Hill2020} represents the brightest example in this SPT protocluster (SPT-PC) survey.

In SPT2349$-$56, most of the observed flux density comes from about 30 galaxies that are all spectroscopically confirmed (by ALMA) to be at redshift 4.3 \citep[][Apostolovski et al.\ in prep.]{Miller2018, Hill2020, Rotermund2020}. If the other fields are similar to SPT2349$-$56, then this catalogue of submm-selected candidate protocluster fields will prove very useful for studying the complex interplay between star formation and large-scale structure formation in the early Universe.

In this paper, we report on sensitive 870-$\mu$m follow-up observations of the 1.3--1.9\,Mpc environment of nine SPT-selected protocluster candidates using the Atacama Pathfinder Experiment (APEX) telescope's Large APEX BOlometer CAmera \citep[LABOCA;][]{kreysa2003,siringo2009}. In Sect.~\ref{observations} we describe the selection criteria used to identify these nine fields in more detail and outline our new LABOCA observations and existing submm data. In Sect.~\ref{analysis} we discuss the data analysis procedures used to identify 870-$\mu$m sources and measure their flux densities. In Sect.~\ref{results} we present the number counts, fractional overdensities, {\it Herschel}-SPIRE colours, and star-formation rates. We compare our results in each section with those from \citet{Lewis2018}, where they analyzed 22 red \textit{Herschel}-SPIRE galaxies. Section~\ref{discussion} discusses these results and the paper concludes in Sect.~\ref{conclusion}. 

The paper assumes a standard $\Lambda$CDM model with cosmological parameters taken from \citet{Planck2018}. In cases where we need to model the IR spectral energy distribution (SED), we apply a modified blackbody function with a dust temperature of 39\,K \citep{Strandet2016}, $\tau_{\text{dust}}=1$ at 100\,$\mu$m, and a dust emissivity index of 2 \citep{Greve2012}, scaled to the redshift of the field.

\section{Observations}
\label{observations}

\subsection{Unlensed sources in the South Pole Telescope millimetre-wave point-source catalogue}

The SPT collaboration has carried out a 2500\,deg$^2$ survey of the sky at 1.4, 2.0, and 3.0\,mm wavelengths. A catalogue of bright mm-wave point sources, unresolved by the SPT's 1-arcmin beam, were identified in this survey \citep{vieira2010, mocanu2013, Everett2020}. Initially, SPT SMGs were selected by their 1.4\,mm raw flux densities, requiring $S_{1.4}\,{>}\,20$\,mJy, and a significance greater than 4.5$\sigma$. APEX-LABOCA and {\it Herschel-\/}SPIRE observations were used to refine the positions of the brightest point sources, and a refined catalogue of 81 sources with $S_{870}\,{>}\,25$\,mJy was subsequently followed up to obtain a complete sample of spectroscopic redshifts \citep{Reuter2020}. 

Given the brightness of the catalogue sources, these objects were likely to be either strongly gravitationally lensed galaxies or collections of distant galaxies, potentially at common redshifts. Gravitational lensing was the most likely explanation for more than 90\,per cent of this sample \citep[e.g.][]{spilker2016}. The remaining 10\,per cent of this bright sample cannot be easily modelled by gravitational lensing \citep[see][]{spilker2016}. However, six sources from this survey stand out by exhibiting multiple ALMA counterparts, where most are unlensed (Chapman et al.\ in prep), with SPT0311$-$58 having the highest magnification factor of $\mu \approx 2$ \citep{spilker2016, Marrone2018}. 

One source, SPT2052$-$56, was not bright enough to be included in the initial point source catalogue of \citet{Everett2020}, but showed significant spatial extent in the LABOCA and SPIRE follow-up observations, and a spectroscopic redshift of $z\,{=}\,4.257$ was secured in $^{12}$CO lines through an ALMA spectral scan (Chapman et al.\ in prep). Similarly, two other sources, SPT0303$-$59 and SPT2018$-$45, were not included in the catalogue of \citet{Reuter2020} as they were $S_{870}\,{<}\,25$\,mJy, but also showed extended 870-$\mu$m structure. While no spectroscopic redshifts are available for these two sources, we can constrain their redshifts photometrically using the pipeline discussed in \citet{Reuter2020} and the photometry presented in Appendix~\ref{Catalogue}.

As candidates for protocluster systems, deeper follow-up observations targeted this sample of nine SPT sources. In the case of one of these sources, SPT2349$-$56, the interpretation as a massive protocluster is unequivocal from extremely detailed studies with ALMA and other facilities \citep{Miller2018, Hill2020}; for the other cases, the evidence is still somewhat insufficient (Chapman et al.\ in prep.). This paper describes one aspect of this follow-up campaign that attempts to characterize these SPT sources, namely an extensive APEX-LABOCA program of 160\,hours to obtain deep, near confusion-limited  870-$\mu$m maps of the environments surrounding these mm sources. These nine SPT targets span a redshift range of $z\,{=}\,$3--7, and represent a density of 0.1\,sources per Gpc$^3$. 

\subsection{APEX-LABOCA observations}

\begin{table}
\caption{Properties of LABOCA fields}
\label{tab: instrument}
\begin{threeparttable}
\begin{tabular}{llllll}
    \hline
    Field & RA & DEC & $t_{\text{int}}^{\,\ast}$ & $d_{\text{cent}}^{\,\dagger}$ & Area \\
    & [J2000] & [J2000] & [hrs] & [mJy] & [arcmin$^2$]\\
    \hline
    \hline
    \input{instrument}
    \hline
\end{tabular}
\begin{tablenotes}
\item[$\ast$] Total integration time of all exposures.
\item[$\dagger$] Central depth of LABOCA fields.
\end{tablenotes}
\end{threeparttable}
\end{table}

The nine SPT-PC fields were first observed at 870-$\mu$m by the APEX telescope's LABOCA instrument  \citep{siringo2009},  as part of a survey of the full sample of SPT sources (project ID M-0101.f-9518C-2018, PI Wei\ss). These shallow maps  (typically 2--4\,mJy r.m.s.) were followed up by deeper targeted LABOCA observations of the nine SPT-PC fields through two additional programmes (ID 0101.A-0475(A), PI Chapman and ID 299.A-5045(A), PI Chapman). These deeper follow-up integrations targeted each of the nine fields for typically 15--20\,hrs, and yielded sensitive maps, covering a combined area of approximately 1300\,arcmin$^{2}$. Observing details are listed in Table~\ref{tab: instrument}.

Our sample was observed over six observing runs from 2018 September to 2019 March. This instrument's passband response is centred on 870\,$\mu$m (345\,GHz) and has a half-transmission width of about 150\,$\mu$m (60\,GHz). Targets were observed in a compact raster-scanning mode, whereby the telescope scans in an Archimedean spiral for 35\,s at four equally spaced raster positions in a $27\,{\rm arcsec}\times27\,{\rm arcsec}$ grid. Each scan was approximately 7\,minutes long, such that each raster position was visited three times, leading to a fully sampled map over the full 11\,arcmin diameter field of view of LABOCA. Each target received an integration time that was on average $t_{\rm int}$\,=\,17\,hr. 

During our observations, we recorded typical precipitable water vapour values between 0.4 and 1.3\,mm, corresponding to a zenith atmospheric opacity of $\tau$\,=\,0.2--0.4. Finally, the flux density scale was determined to an r.m.s. accuracy of 7\,per cent, using observations of the primary calibrators Uranus, and Neptune, while pointing was checked every hour using nearby quasars and found to be stable to an r.m.s. of 3\,arcsec.

\subsection{\textit{Herschel}-SPIRE observations}

We also use \textit{Herschel}-SPIRE \citep{Griffin2010} data at 250, 350, and 500-$\mu$m, primarily obtained during the initial multiwavelength follow-up campaign. Specifically, the original follow-up proposal (programme ID OT2\_jvieira\_5, PI Vieira) targeted seven of the nine protocluster candidate fields described above, with a total area of approximately 2000\,arcmin$^{2}$. Separately, SPT0311$-$58 was observed as part of a director's discretionary time proposal (programme ID DDT\_mstrande\_1, PI Strandet), while SPT2335$-$53 falls in the large \textit{Herschel}-SPIRE 90\,deg$^2$ survey already carried out to complement the SPT data \citep[see][]{holder2013}.

\section{Data analysis}
\label{analysis}

\subsection{LABOCA source extraction}

After processing the data using the standard BOlometer Array Analysis Software \citep[{\tt BOA};][]{Schuller2012}, we find that our maps reach central depths of 1.0--1.5\,mJy. We convolve the LABOCA flux density and noise maps with a 18.6\,arcsec Gaussian beam in order to produce maximum-likelihood signal-to-noise-ratio (SNR) maps for point-source detections \citep[see e.g.,][]{scott2002,coppin2006}, see Fig.~\ref{fig: 0303}. Following the threshold adopted in \citet{Weiss2009}, a detection threshold of 3.7$\sigma$ was chosen; we find that at this threshold, the total number of negative peaks is 5\,per cent of the total number of positive peaks (this being an estimate of the false detection rate). We fit Gaussians to the brightest peaks in the beam-convolved maps and subtract them until all the remaining peaks are less than our threshold of 3.7$\sigma$ to measure flux densities with corresponding uncertainties. We chose this source detection method because the LABOCA beam resolves the central core in most fields. By deconvolving the bright and extended LABOCA central cores (and the surrounding sources), we can identify multiple sources in the resolved emission (between two to eight sources for each SPT field, with SPT0303$-$59 showing the most extended, blended central complex). In comparison, the sum of the flux densities of the sources identified in the resolved emission is comparable to the flux density found by applying aperture photometry to the same region. 

\begin{figure}
    \includegraphics[width = \twidth]{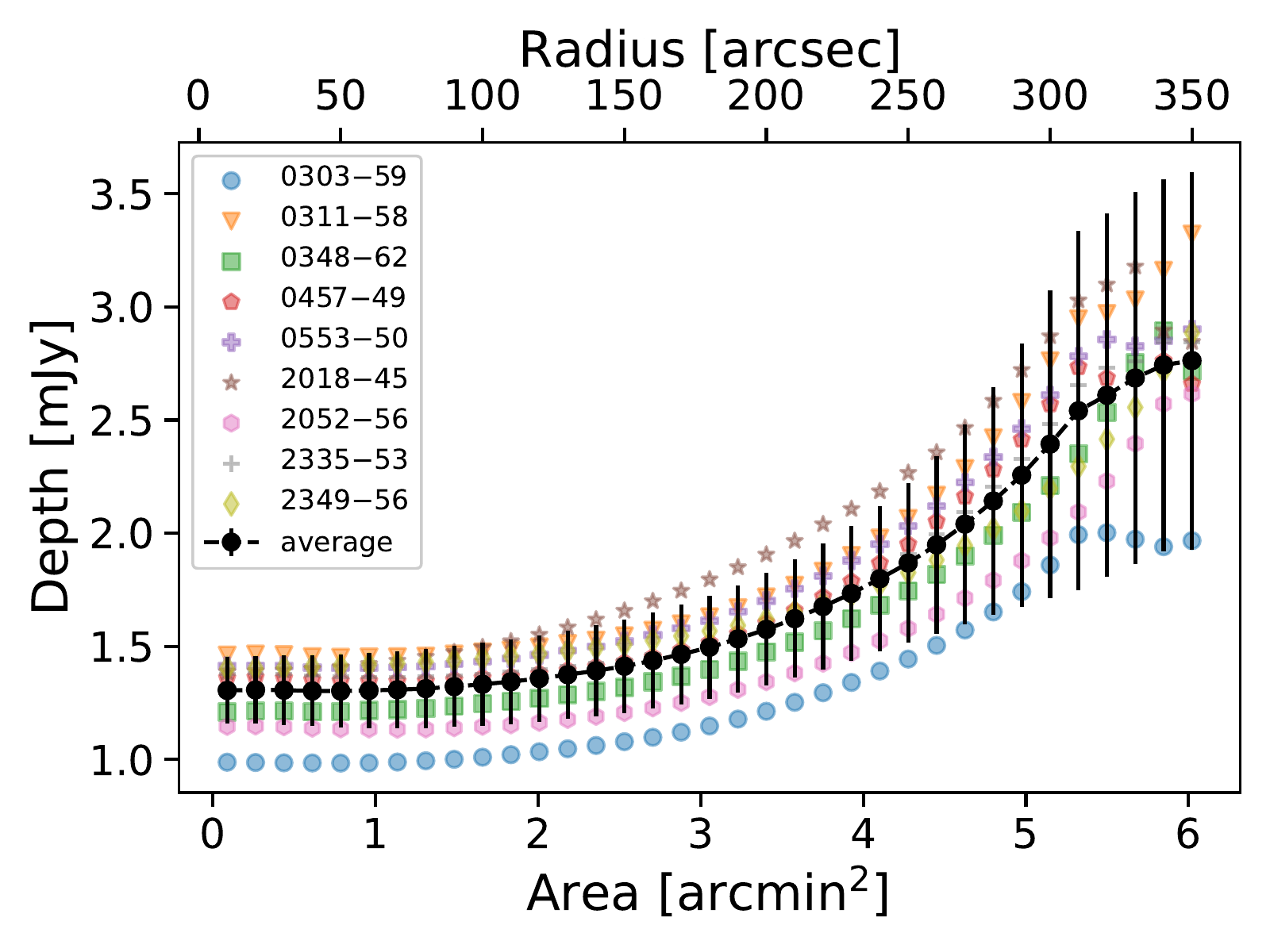}
    \caption{Depth of the average and individual SPT fields, calculated between concentric circular annuli. We show the radius of the outer annuli and the area of the annuli. The depth is relatively uniform, with only a 10\,per cent increase at 160\,arcsec and increasing by 50\,per cent at 240\,arcsec; this suggests that source detection is complete in the inner regions.}
\label{fig:rms}
\end{figure}

We plot the depth of the LABOCA fields as a function of both area and radius of concentric circular annuli (see Fig.~\ref{fig:rms}) and show that they vary approximately 50\,per cent between the centre and 240\,arcsec. At 350\,arcsec, the average depth increases to 2.8\,mJy, which is about 2.1 times higher than the central depth. The relatively slight increase in depth suggests that a uniform detection threshold of 3.7$\sigma$ is sufficient, and only at the outer edges of the LABOCA fields are we incomplete in our source catalogue.

For this paper, we identify the sources with the highest SNRs (16.8--41.5) in each field as ``central sources'' (labelled A in Appendix~\ref{Catalogue}). We note that some LABOCA 870-$\mu$m flux densities were previously derived in \citet{spilker2016}, and \citet{ Strandet2016} using older and shallower data. We compare our newly-derived flux density measurements to these older values and found that the two agree within the uncertainties, with no evidence for any systematic shifts. 

Using the significance threshold and source definition outlined above, we identify 98 sources across the nine fields. We provide the flux densities for these sources in Appendix~\ref{Catalogue}. We clarify that the sources presented for SPT2349$-$56 do not share the same labels as those found in \citet{Miller2018}, and \citet{Hill2020} (where the sources resolve into many individual galaxies).

\subsection{LABOCA flux deboosting}

Since our source-extraction technique involves a SNR cut, there is a systematic boosting of lower flux objects due to Eddington-type bias. Even though the objects will be normally distributed under our assumption of Gaussian statistics, the underlying population distribution of luminous objects means that intrinsically, there are many more faint objects in our fields below our detection threshold than bright objects above. We discard the ones below the threshold and on average, the flux density of an object near the threshold will be overestimated. 

Our raw flux densities need to be ``deboosted'' to correct for this effect. We use the method described in \citet{coppin2005, coppin2006}, where we combine the assumed Gaussian likelihood distribution of a source, having a mean flux density $S_{\nu}$ and uncertainty of $\delta S_{\nu}$, with a prior given by the extragalactic differential number counts \citep[as estimated from LABOCA data by][]{Weiss2009} to obtain the posterior distribution for the actual flux density. We take the posterior distribution peak to be the deboosted flux density of each source, with the uncertainties determined by calculating 68\,per cent confidence intervals. The tables providing these deboosted flux densities are in Appendix~\ref{Catalogue} (where they are labelled $S_{870}^{\rm deb}$).

For sources near our detection threshold of 3.7$\sigma$, deboosting will have the most substantial effect on their flux densities. The posterior distribution for some of these sources peaks at the lower flux limit of the background prior, and so we only provide upper limits for their flux densities (we quote the 99.7\,per cent confidence intervals). Of the 98 identified sources, 36 of them have only upper limits on their 870-$\mu$m flux densities.

\subsection{\textit{Herschel}-SPIRE photometry}

Since our \textit{Herschel}-SPIRE maps are significantly confused by the cosmic infrared background (i.e. sources near to the line of sight at redshifts other than that of the SPT source), we need to use an appropriate filter to measure the flux densities of our sources at 250, 350, and 500-$\mu$m. Convolving these maps with the point-spread function (PSF) is sufficient for isolated point-source detection of well-characterized data, but in this case, an optimal filter must take into account the background source number counts. We choose to filter all of our SPIRE maps using the matched-filter technique described in Appendix~A of \citet{Chapin2011}, where, in Fourier space, the filter is the PSF weighted by the sum of the noise variance terms. In this case, the ``noise'' variance is a combination of instrumental white noise (estimated as the map's r.m.s.), plus Poisson noise from sources in the background \citep[taken from][]{Nguyen2010}. After we convolve the raw SPIRE maps with these filters, it is still the case that some of the central cores are extended in the 17.6--35.2\,arcsec SPIRE beam. Gaussian deconvolution of the core results in an overestimate of the flux densities for such resolved objects, so instead, we measure the aperture photometry of the entire central core. We apply a set of corrections following the prescription outlined in the SPIRE Data Reduction Guide\footnote{\url{http://herschel.esac.esa.int/hcss-doc-15.0}} and the resulting flux density measurements are given in Appendix~\ref{Catalogue}.

We note that SPIRE flux densities at 250, 350, and 500-$\mu$m for the central sources in some of these fields were available in \citet{spilker2016}, \citet{Strandet2016}, and \citet{Reuter2020}; there they were derived simply by convolving the raw maps with the PSF and measuring the values of the peak pixels, while our approach takes into account the effects of confusion. Upon comparing these two flux density estimation methods, we find the results are consistent within the considerable uncertainties, and we see no evidence of systematic differences.

\begin{figure*}
    \includegraphics[width = \twidth]{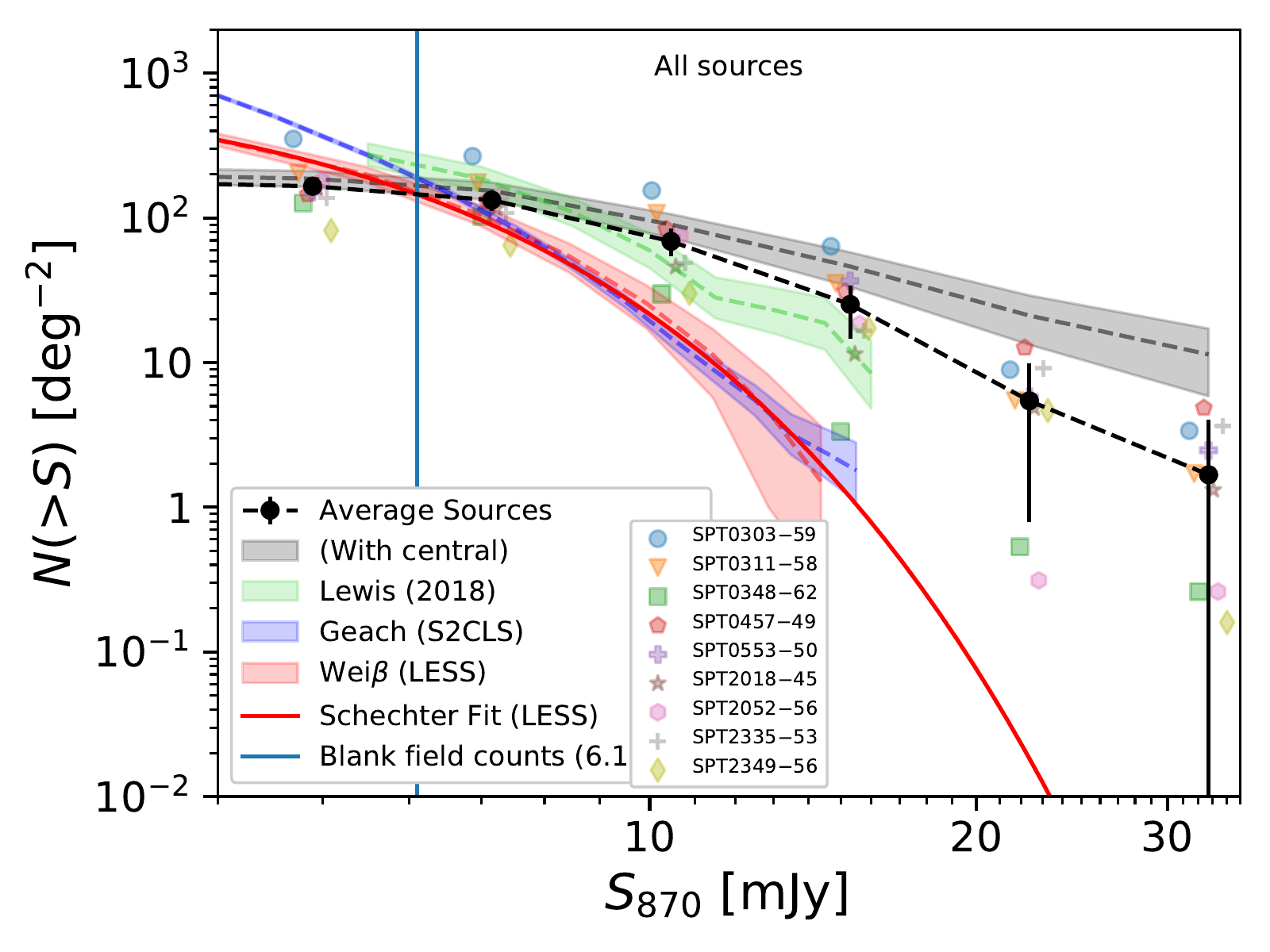}
    \includegraphics[width = \twidth]{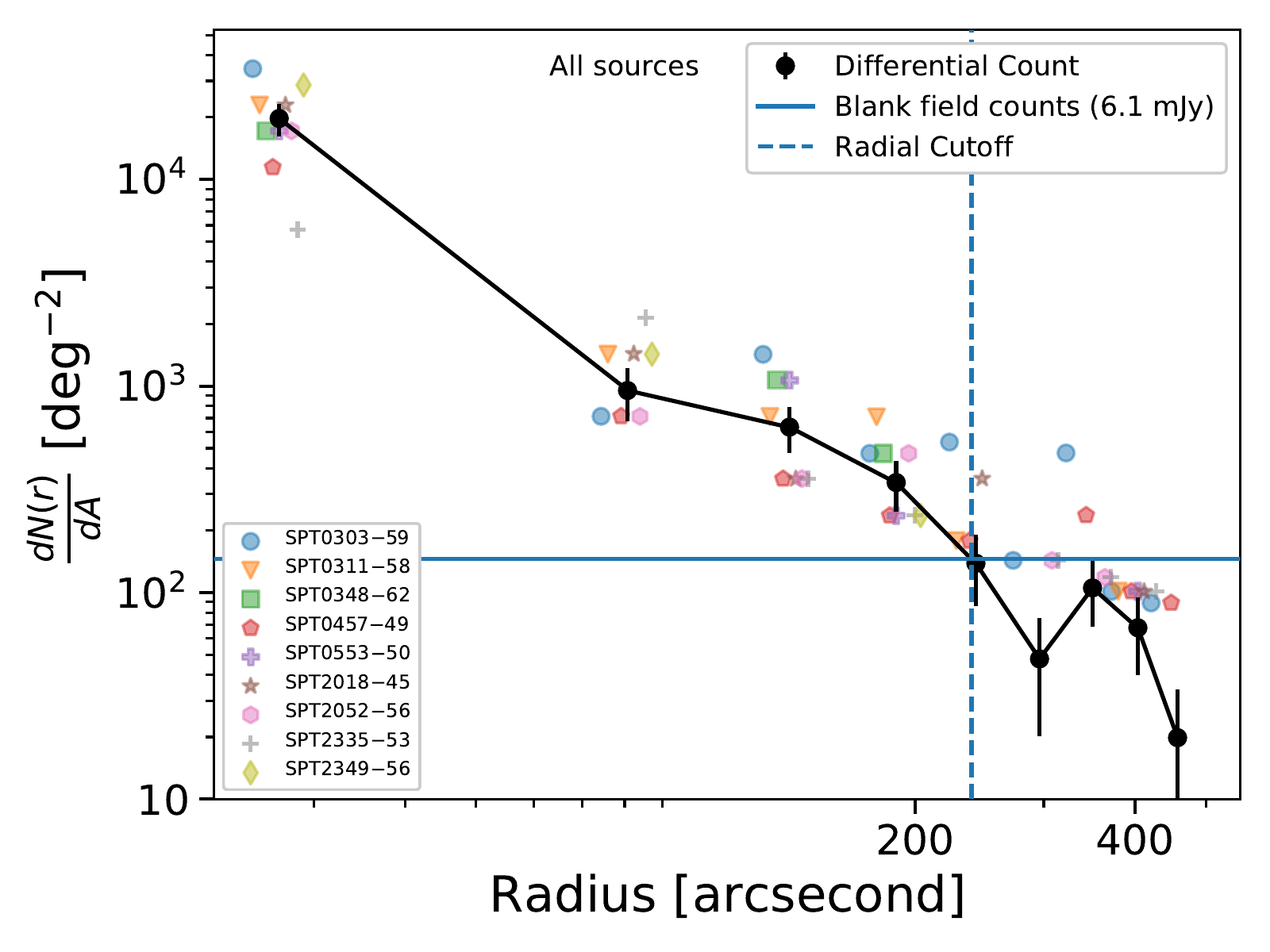}
    \caption{\textit{Left}:  Cumulative number counts of all sources before applying a radial cutoff for individual fields (coloured symbols), averaged over all fields excluding the nine central sources (black, 89 sources) and including the nine central sources (shaded grey region, 98 sources). The errors are the quadrature sum of the Poisson noise and 68\,per cent confidence range. We also show the cumulative number counts from LESS \citep[][red shaded region, with a Schechter function fit to the data as a red line]{Weiss2009} and S2CLS \citep[][blue shaded region]{Geach2017}, which are both large ``blank sky'' surveys of submillimetre galaxies (SMGs). The cumulative number counts from 86 SMGs selected by 22 red \textit{Herschel}-SPIRE galaxies is also plotted \citep[][green shaded region]{Lewis2018}. The blue vertical line shows where our number counts intersect with the background number counts from LESS, at a value of about 130\,deg$^{-2}$, which we use to define our radial cutoff. \textit{Right}: Differential radial number counts for all 98 LABOCA-detected sources.  These are shown field by field (coloured symbols) and averaged over all fields (solid black), where the radius is the distance from the target source with corresponding Poisson noise as the error. The horizontal blue line shows the value at which our total cumulative number counts intersect with the background counts from \citet{Weiss2009} (around 130\,deg$^{-2}$). This line intersects with the radial distribution at about 240\,arcsec; beyond this radius, we expect to be statistically detecting mostly background SMGs, and thus we define this to be our radial cutoff for analyzing candidate sources. The radial density is several times higher than the blank-field counts around the target field and is evidence that these fields might be protoclusters.}
\label{fig:radialCut}
\end{figure*}

\section{Results}
\label{results}

\subsection{Radial counts analysis}

For all of our LABOCA fields, we compute the number counts cumulative in flux density and differential counts as a function of distance from the central sources. We estimate the uncertainties in flux density by performing a Monte Carlo simulation, where we draw flux densities from each source's posterior distributions. Figure~\ref{fig:radialCut} shows the resulting cumulative flux and radial distributions; here, we have taken the maximum likelihood and 68\,per cent confidence intervals from the Monte Carlo simulations. We show the cumulative counts with and without the nine central sources' inclusion since they might bias the number counts. The cumulative flux distribution is compared to the blank field 870-$\mu$m cumulative number counts from the LABOCA Extended Chandra Deep Field South (ECDFS) Submillimetre Survey \citep[LESS: 0.25 deg$^2$,][]{Weiss2009} and the Submillimetre Common-User Bolometer Array 2 (SCUBA--2) Cosmology Legacy Survey \citep[S2CLS: 5 deg$^2$,][]{Geach2017}. We also compare our cumulative flux density distribution to that of \citet{Lewis2018}, who performed a similar LABOCA follow-up study of 22 red {\it Herschel}-SPIRE objects identified in several surveys. The lack of bright sources is due to removing all $S_{870}>15$\,mJy sources without lensing models. Our number counts intersect with the blank field counts at 130\,deg$^{-2}$ with a flux density of 6.1\,mJy, which implies that our sample is complete down to this level. The constant depth within the inner region of 130\,deg$^{-2}$ and the average median flux density of the dimmest source being at 6.0\,mJy allows us to believe that we are complete down to this flux density.

Turning to our radial distributions, we can see that the central cores are more overdense than the surrounding regions, and beyond roughly 240\,arcsec the number counts reach the blank-field counts of 130\,deg$^{-2}$. We also observe that the density is several times higher than the blank field count measured at the intersection of our average number counts with the background counts \citep{Weiss2009}. In the analysis that follows, we will adopt 240\,arcsec as a radial cutoff where we see an excess of sources surrounding the central sources and are not part of the overall structure and we also assume all sources within the radial cutoff are part of the structure, without any interlopers or false detections. This cutoff also corresponds to a 50\,per cent increase in the maps' r.m.s. values (see Fig.~\ref{fig:rms}). This cut removes 25 sources from our catalogue, with 57 of the remaining sources being both within the radial cutoff and not deboosted down to the survey limit (i.e. they only have upper limits). Appropriate headers in Appendix~\ref{Catalogue} separate the categories of sources in each field where we present the associated flux densities at 250, 350, 500, and 870-$\mu$m.

\subsection{Number counts within 240 arcsec}

\begin{figure*}
    \includegraphics[width = \twidth{}]{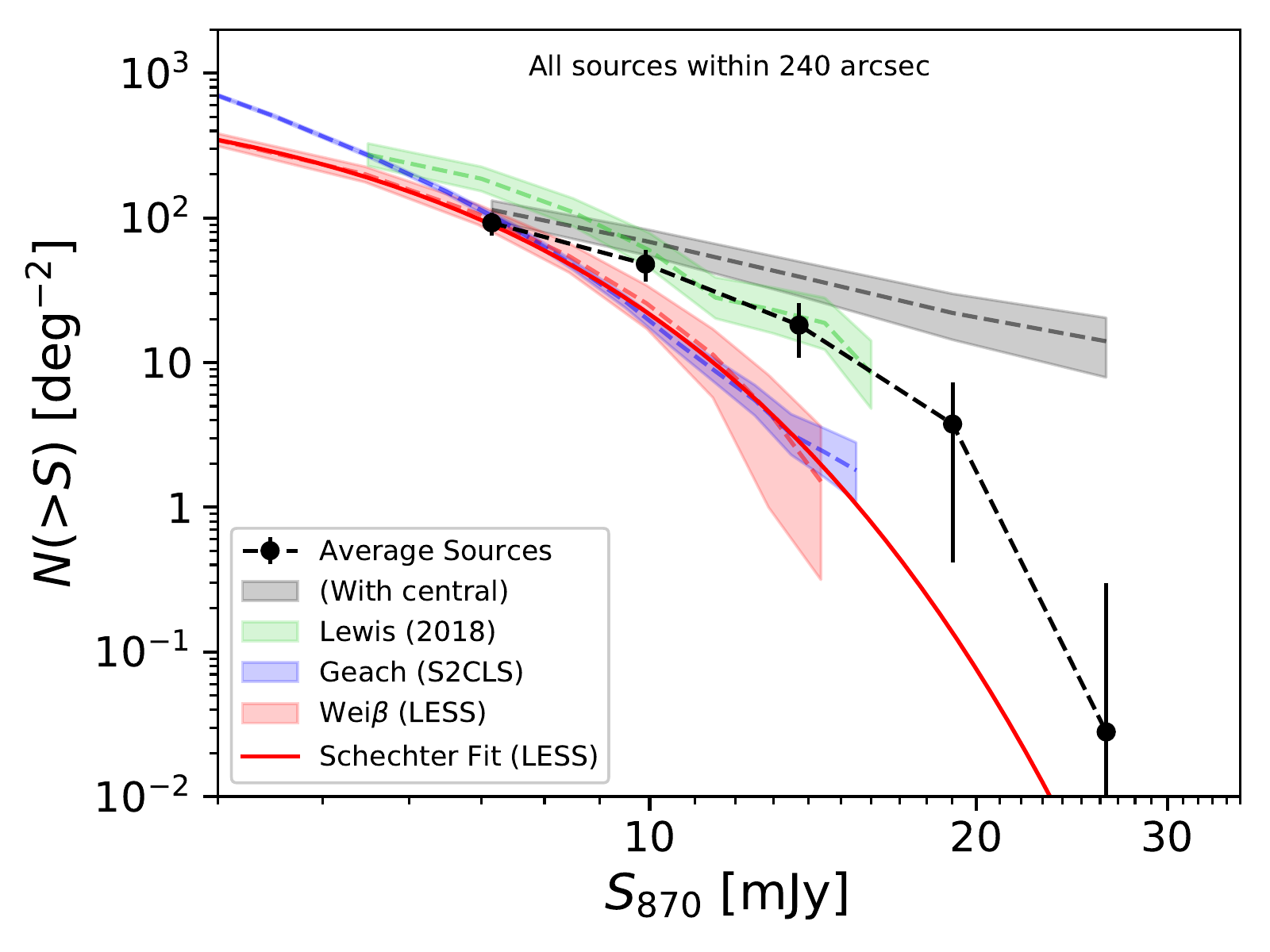}
    \includegraphics[width = \twidth{}]{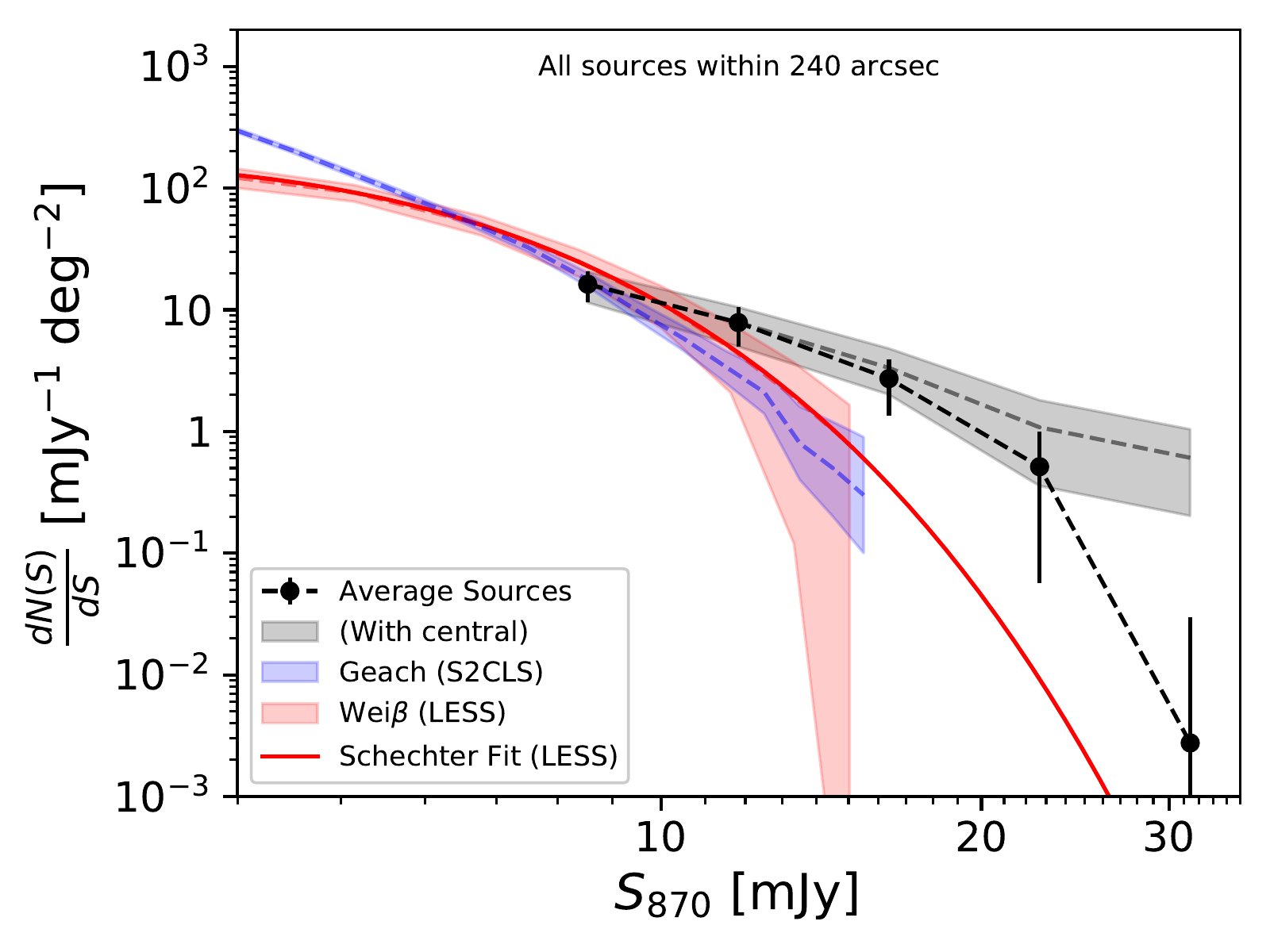}
    \caption{\textit{Left}: Cumulative number counts of the 57 LABOCA sources remaining after removing the sources outside the 240\,arcsec radial cutoff (shaded grey) and the nine central sources removed (black). In this plot we again show the background number counts from LESS \citep[][red shaded region]{Weiss2009}, S2CLS \citep[][blue shaded region]{Geach2017}, and \citet{Lewis2018} (green shaded region). The solid red line shows the best-fit Schechter function for the LESS number counts. \textit{Right}: Same as the left panel, but now showing the differential number counts in five flux-density bins. The number counts of these SPT fields are above the background and show strong similarities to the counts of \citet{Lewis2018} at higher flux densities.} 
    \label{fig:counts}
\end{figure*}

In Fig.~\ref{fig:counts}, we show the total cumulative and differential number counts across all fields, after removing sources outside our radial cutoff. Comparing the number counts to those from LESS \citep{Weiss2009} and S2CLS \citep{Geach2017}, a clear enhancement can be seen in the number of bright ($S_{870}\,{\gtrsim}\,10\,$mJy) sources. The counts extend to a much higher flux density compared to the ``blank sky'' surveys, and even after extrapolating such surveys to higher flux densities (e.g. by fitting a Schechter function to the background counts), we still find that our source counts are about an order of magnitude higher.

The cumulative distribution is compared with the distribution calculated from the 86 SMGs detected by LABOCA in \citet{Lewis2018}, who also removed the bright central sources, as was done here (but did not provide differential counts). The two number counts are comparable within their flux density limits.

\subsection{Fractional overdensity}

\begin{figure}
    \includegraphics[width = \twidth{}]{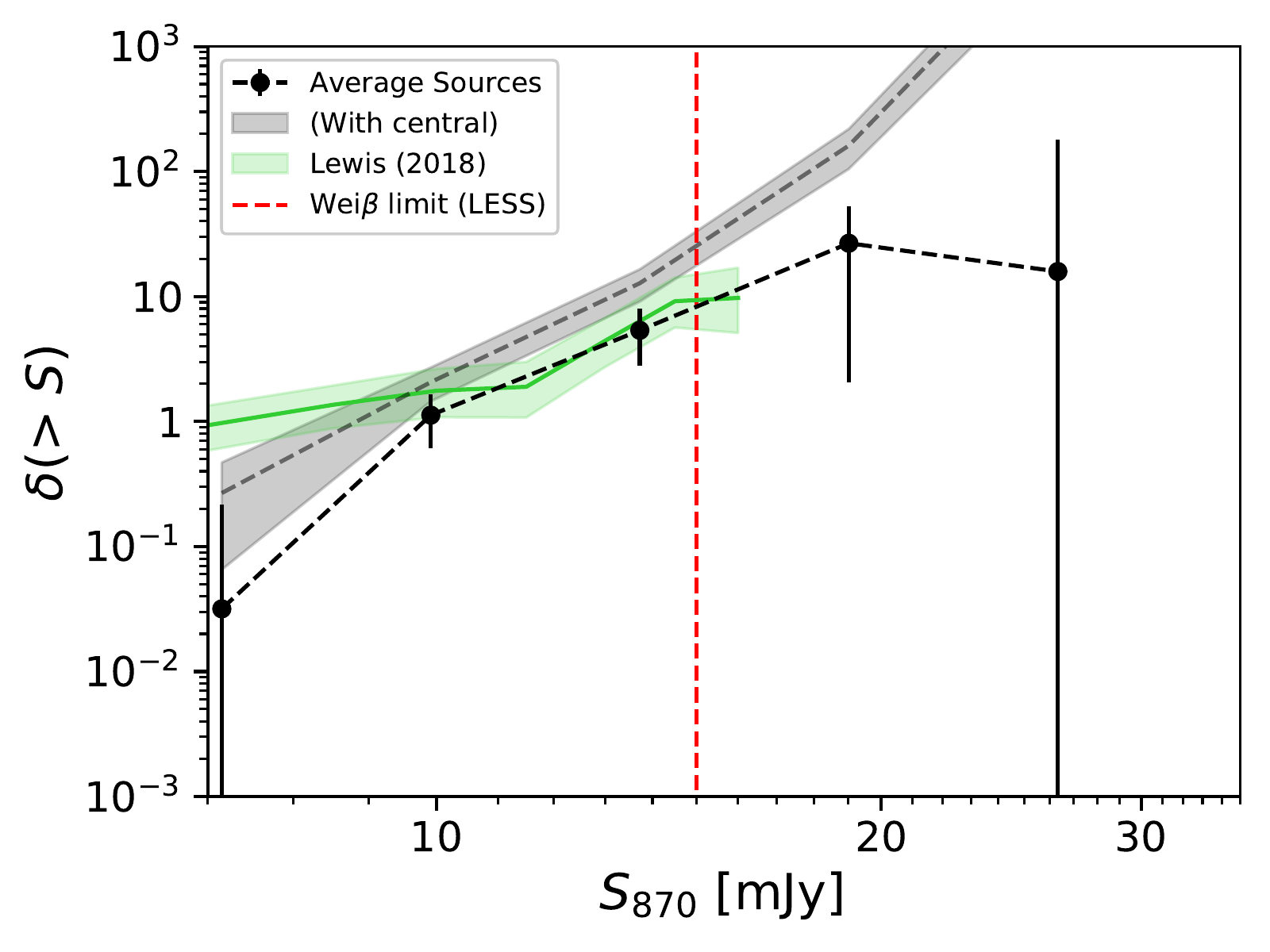}
    \caption{Overdensity of LABOCA-detected sources compared to background sources \citep[Schecther fit of LESS,][]{Weiss2009} as a function of the 870-$\mu$m flux densities in grey and with the central sources removed in black. The indicated errors are from the fields' cumulative counts. If we were to include the Poisson error of the background counts, then above 15\,mJy the lower limit of the error will reach into the underdense regime (i.e. the estimate will be completely uncertain). This uncertainty is due to a deficiency of bright sources in LESS above 15\,mJy, which required us to extrapolate using a Schechter function. However, since a lack of bright sources in random fields means that we cannot accurately determine the background counts, there is no doubt that our catalogue contains more bright sources than average fields. The green line and shaded region represent the overdensity of the \citet{Lewis2018} catalogue. The dashed red line represents the number counts limit of LESS. The significant overdensity suggests a physical association among many of the sources we detect in each field.}
\label{fig:overD}
\end{figure}

In Fig.~\ref{fig:overD}, we show the fractional overdensities of these fields as a function of flux density, computed by taking the ratio of our cumulative number counts to that from the Schechter fit of the LESS survey and removing the average background value. Since the LESS data are deeper than our LABOCA maps, our counts at fainter flux densities are lower. The LABOCA survey covers 0.36\,deg$^2$ and is comparable to the 0.25\,deg$^2$ of the Chandra Deep Field South, where the excess bright sources are attributed to selection effects. We note that the brightest source found in LESS is around 15\,mJy; thus, we extrapolate the overdensity beyond this limit with a Schechter function. Moreover, at higher flux densities, the number of sources in the LESS survey decreases dramatically, leading to more considerable uncertainties. The overdensities of our catalogue and those from \citet{Lewis2018} are comparable in the range 8--16\,mJy. Overall, our target fields contain considerably more bright sources than the average part of the sky ($\delta\,{\approx}\,$10 at $S_{870}\,{=}\,14$\,mJy).

\subsection{SPIRE colours}

In Fig.~\ref{fig: 0303}, alongside our LABOCA images, we show $S_{870}/S_{350}$ colours as a function of $S_{870}$ for all of the LABOCA sources detected in the protocluster candidate fields (after removing those sources outside the radial cutoff). In these plots, we can compare the colours of our sources (all sources within the 240\,arcsec cutoff and show the integrated colour, and flux for the extended central core(s)) to field SMGs from \citet{Swinbank2014}. The grey horizontal lines and shaded regions show the averages and standard deviations of the field SMGs, while the red horizontal lines and shaded regions are the averages and standard deviations computed field by field.

In each field, we find that our LABOCA-detected sources' average colours are redder than typical SMGs, although small-number statistics in several fields means that the uncertainties are quite large. This behaviour is expected if our sources are SMGs at higher redshift than the far-IR (FIR) ``foreground,'' which is typically at $z\,{\lesssim}\,2$ \citep{Marsden2009, Wynne2012}. Therefore, the sources we have found in these fields are consistent with the central sources instead of random foreground galaxies. We excluded photometric redshifts for individual sources from our analysis due to the uncertainties in our \textit{Herschel}-SPIRE photometry. 

\subsection{Star-formation rates}

The mm/submm brightness of a high-$z$ galaxy is closely linked to its SFR because dust tends to enshroud star-formation. It absorbs starlight and thermally re-radiates it at FIR wavelengths and is subsequently redshifted into the mm/submm regime. By fitting FIR photometry to a template SED (typically a modified blackbody), the total FIR luminosity, $L_{\rm FIR}$, found by integrating from 42 to 500-$\mu$m, can be converted to an SFR estimate using a scaling relation of the form ${\rm SFR}[{\rm M}_{\odot}\,{\rm yr}^{-1}]\,{=}\, 0.95\,{\times}\,10^{-10}\,{\rm L}_{\rm FIR}[{\rm L}_\odot$] \citep{Kennicutt1998}.

Our \textit{Herschel} photometry is quite confused and uncertain, and thus we only use our LABOCA 870-$\mu$m photometry here to scale a modified blackbody function with a dust temperature of 39\,K (the mean value for lensed, individual SPT sources; \citealt{Strandet2016}) and a dust emissivity index of 2 \citep{Greve2012}. Integrating the function gives the total FIR luminosity, which we convert to an SFR using the relation above (see Table~\ref{tab:SFR}). Comparing the SFRs calculated by the FIR integration and the relation from \citet{dudzeviciute2020} (where they fit their IR data to a model), we recognize the SFRs from FIR integration are roughly 3.5 times higher than the ones calculated from the relation. The difference may be attributed to the mapping of FIR dust emission with a single photometry point, but the difference can only be determined if we perform the same modelling \citep{dudzeviciute2020} for our sources. As a rough estimate of the systematic uncertainty, a $\pm$\,5\,K change in temperature leads to average variations of 30--40\,per cent in our estimate of SFR. If we assume a $T_{\text{dust}}$ 52.4\,K \citep{Reuter2020}, our SFRs will be approximately two times as large. For the rest of the paper, we quote the integrated SFRs to align with other studies \citep[e.g.][]{Chapman2010, Casey2016, Lewis2018, Hill2020, Reuter2020}.

In Table~\ref{tab:SFR}, we provide both the SFRs of the central sources and the SFRs integrated over the whole field. The surrounding LABOCA sources might also not be all at the same redshifts \citep{Hayward2013, Cowley2015, Hayward2018, Wardlow2018}, which would result in the wrong SFR estimates. Hence, we provide maximum and minimum SFRs given the assumption that we include either all or none of the LABOCA sources.

\begin{figure}
    \includegraphics[width = \twidth{}]{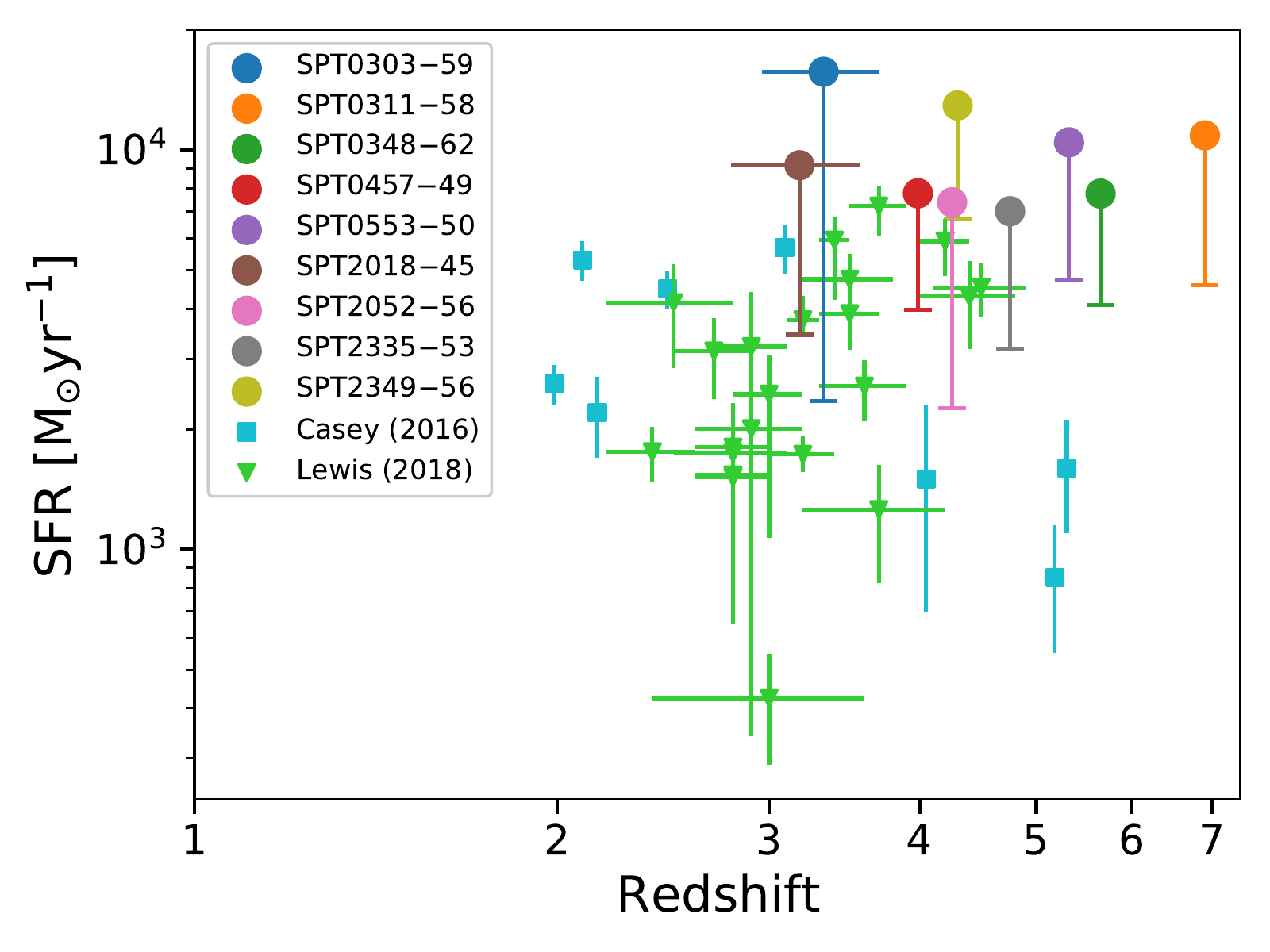}
    \caption{Maximum SFRs of each field (i.e. adding all LABOCA source contributions) are shown as a function of redshift. The SFRs are derived here by scaling a modified blackbody function by the total 870-$\mu$m flux density, integrating from 42 to 500-$\mu$m, and multiplying by a proportionality factor of $0.95\,{\times}\,10^{-10}\,{\rm M}_{\odot}\,{\rm yr}^{-1}{\rm L}_{\odot}^{-1}$ \citep{Kennicutt1998}. The redshifts of our sample are spectroscopic when available \citet{Reuter2020} and otherwise are calculated from photometry. The blue squares are protoclusters in the literature taken from \citet{Casey2016}, while the triangles are from \citet{Lewis2018} where they use the Kennicutt relation to calculate the SFRs. We also show the minimum SFRs with only the central sources as lower limits. The horizontal errors show the uncertainties in the photometric redshifts. The SFRs of the nine fields are higher (or at the high end if we only consider the central sources) in comparison to these protoclusters.}
\label{fig: SFR}
\end{figure}

In Fig.~\ref{fig: SFR}, we show the minimum and maximum SFRs of each field as a function of redshift. In this plot, we also show the total SFRs of other protocluster fields from the literature (where they include their central sources; see \citealt{Casey2016} and \citealt{Lewis2018} for details).  The SPT fields have star formation rates higher than those seen in other protocluster samples at $z$\,$>$\,3 assuming that most of the brightest LABOCA sources are confirmed to lie at the same redshifts.

\section{Discussion}
\label{discussion}

The combination of strong fractional overdensities and compacted central region (240\,arcsec radius) exhibited in these nine fields suggests that these sources might correspond to coalescing structures that will become some of the most massive galaxy clusters in today's Universe. However, the lack of redshift data on these LABOCA sources makes it difficult to separate the interlopers from the actual members of the high-$z$ structures, and we will still find a dramatic excess in overdensities if we were to subtract out field contributions. Therefore, we classify these clusters as protocluster candidates. Individually determining which sources are members of the same structure requires additional spectroscopy \citep[e.g.,][]{Hayward2018}; the exception to this is SPT2349$-$56, where spectroscopy has already confirmed 23 galaxies in the central core alone, and classify it as one of the brightest and highest redshift protoclusters known \citep[see][]{Miller2018, Hill2020}.

Because 870-$\mu$m flux density is linked to star formation, we find that these candidate protocluster fields with bright submm galaxies are undergoing an active phase in their formation; higher SFRs than seen in typical candidate protoclusters described in the literature \citep[e.g.,][]{Casey2016}. Recall that the candidate protocluster fields presented in this paper were initially selected due to their bright mm flux densities, while many other protocluster fields were selected from the overdensities of galaxies that are bright in the near-IR or through blind redshift surveys in the optical. In this context, the candidate protocluster fields described here represent an epoch in cluster formation where star formation is at its maximum level.

These mm/submm-selected protocluster candidates tend to have higher redshifts than other protoclusters, which may be related to downsizing, where larger overdensities form their stars and subsequently quench earlier than smaller objects \citep[e.g.,][]{magliocchetti2013,miller2015,wilkinson2017}. This characteristic would mean that our sample of protoclusters candidates is probing an early epoch of the largest and rarest galaxy clusters seen today. Recent work analyzing the MultiDark-Planck-2 \citep[MDPL2;][]{riebe2013,klypin2016} simulation for the $z\,{=}\,$0 counterparts of $z\,{\simeq}\,4$ mergers of massive dark matter halos suggests that, if we see such massive mergers in these fields, these structures will grow to be the most massive clusters, of order 10$^{15}$\,M$_{\odot}$ today \citep{Rennehan2019}.

To provide some useful metric with which future simulations can compare, we have computed the maximum SFR volume density of each of these fields and present our estimates in Table~\ref{tab:SFR}. Here we have estimated the volumes by simply assuming spherical symmetry around the central sources, and we have taken the radius containing the maximum SFR to be 240\,arcsec (1.3--1.9\,Mpc).

In Table~\ref{tab:SFR} we can see that the maximum SFRs of these fields reach values 7000--16000\,${\rm M}_{\odot}\,{\rm yr}^{-1}$ with corresponding volume densities of several hundred ${\rm M}_{\odot}\,{\rm yr}^{-1}\,{\rm Mpc}^{-3}$. Note that these values are biased slightly high, since we have not subtracted the field galaxy SFR; based on the background number counts, we expect about two field galaxies per 240\,arcsec aperture, contributing an average of about 1000\,M$_{\odot}$\,yr$^{-1}$, which is a fraction ($<$\,15\,per cent) of the total SFR. These total SFRs are roughly an order of magnitude higher than what current simulations of high redshift protoclusters \citep[e.g.,][]{saro2009,granato2015} show and could be due to the rarity of our candidate protocluster fields. For example, if events such as these occur only once per 10\,Gpc$^3$ \citep{Rennehan2019}, current simulations may not be probing large enough volumes. Additionally, current hydrodynamical simulations of protoclusters may not accurately capture the physics of star formation in these extreme environments due to the necessity of implementing sub-grid models (star formation, stellar feedback, and active galactic nuclei feedback) or they may have insufficient resolution due to the computational expense of simulating such massive halos (Lim et al.\ in prep.).

\begin{table*}
\caption{Flux densities and SFR density estimates.}
\label{tab:SFR}
\begin{threeparttable}
\begin{tabular}{lllllllll}
    \hline
    Field & RA & DEC & $z$ &  \centered{$S_{1.4}^{\text{deb}}$} & \centered{$S_{870}^{\text{int}}$\tnote{1}} & \centered{Max SFR\tnote{2}} & \centered{Min SFR\tnote{3}} & \centered{SFR volume Density\tnote{4}}\\
    & [J2000] & [J2000] & & \centered{[mJy]} & \centered{[mJy]} & \multicolumn{2}{c}{[${\rm 10^{3}M}_{\odot}\,{\rm yr}^{-1}$]} & [${\rm 10^{2}M}_{\odot}\,{\rm yr}^{-1}\,{\rm Mpc}^{-3}$]\\
    \hline
    \hline
    \input{SFRlatex}
    \hline
\end{tabular}
\begin{tablenotes}
\item[deb] Deboosted 1.4\,mm flux densities
\item[1] Integrated flux densities of central core of each field (denoted by integer superscripts in Appendix~\ref{fig: 0303})
\item[2] Total SFR estimated for all sources within 240 arcsec of the centre.
\item[3] SFR of only source A (central source) labelled in Appendix~\ref{Catalogue}.
\item[4] Density within spherical volume of radius of 240\,arcsec.
\item[5] Photometric redshift.

\end{tablenotes}
\end{threeparttable}
\end{table*}

\section{Conclusions}
\label{conclusion}

In this paper, we have reported observations of nine $z\,{=}\,$3--7 protocluster candidate fields at 870-$\mu$m using the LABOCA instrument mounted on the APEX telescope. These fields were discovered in the 2500\,deg$^{2}$ SPT survey, and selected due to their bright flux densities and point-source nature as seen with SPT's 1-arcmin beam. Subsequent follow-up observations using ground- and space-based facilities have provided the resolution necessary to resolve each target field in a bright core and with extended structure. 

Our 870-$\mu$m LABOCA maps reach depths between 1.0 and 1.5\,mJy, and we found 98 sources with a 3.7$\sigma$ cutoff. We measured 870-$\mu$m flux densities and corrected these measurements for the statistical effect of flux boosting. Then we compared the resulting number counts to counts of background SMGs at the same wavelength. We found that beyond about 240\,arcsec, our number counts reach background levels, meaning that 25 sources we found beyond this radius are statistically likely to be field SMGs and thus were removed from our sample of candidate protocluster members.

Using existing \textit{Herschel}-SPIRE data in these nine fields, we measured the 250-, 350-, and 500-$\mu$m flux densities of our LABOCA-detected sources. We computed the mean $S_{870}/S_{350}$ colour of each field and compared these to samples of background SMGs. From this comparison, we saw that our sources are redder than field SMGs, even with the large uncertainties associated with the photometry.

We computed cumulative and differential number counts of our final catalogue of protocluster candidates and used these to derive the fractional overdensity compared to the background sky. Beyond about 10\,mJy, our fields are considerably overdense, reaching ten times the density at 14\,mJy compared to average parts of the sky. Since overdensities of late-type galaxies is an indicator of the seeds of present-day clusters, we classify these nine fields as candidate protoclusters \citep[except for SPT2349$-$56, which has already been confirmed as a protocluster; see][]{Miller2018, Hill2020}, where confirmation requires spectroscopic data.

We also derived SFRs by scaling a modified blackbody SED template to our measured 870-mJy flux densities and compared these to other protocluster fields from the literature. These nine fields contain considerably more star formation than seen in many previously reported protoclusters, likely due to their mm-wavelength selection and unprecedented survey area. Current simulations are unable to achieve the intense SFR densities that we see in our sample.

The development of mm and submm astronomy over the past several decades has led to a substantial amount of observational data relating to star formation and structure formation in the early Universe. Some of the most exciting sources detected in submm surveys are protoclusters, such as those presented here. Simulating such objects is exceptionally challenging due to their rarity and the limited resolution possible when running hydrodynamical simulations of very massive halos. As computational resources grow and codes are made more efficient, such simulations may become possible, and comparing with protocluster observations may lead to valuable insights into the physics of galaxy formation in extreme environments. Additionally, increasing the sample size of protocluster fields such as those described here, through upcoming extensive surveys, will help overcome small-number statistics. This will allow for abundances to be more accurately estimated and enable more precise comparisons with next-generation simulations.

\section{Acknowledgements}
\label{acknowledgements}
This publication is based on data acquired with the Atacama Pathfinder Experiment (APEX). APEX is a collaboration between the Max-Planck-Institut für Radioastronomie, the European Southern Observatory, and the Onsala Space Observatory. The SPT is supported by the National Science Foundation through grant PLR-1248097, with partial support through PHY-1125897, the Kavli  Foundation, and the Gordon and Betty Moore Foundation grant GBMF 947. \textit{Herschel} is an ESA space observatory with science instruments provided by European-led Principal Investigator consortia and with important participation from NASA. SPIRE has been developed by a consortium of institutes led by Cardiff University (UK) and including: Univ. Lethbridge (Canada); NAOC (China); CEA, LAM (France); IFSI, Univ. Padua (Italy); IAC (Spain); Stockholm Observatory (Sweden); Imperial College London, RAL, UCL-MSSL, UKATC, Univ. Sussex (UK); and Caltech, JPL, NHSC, Univ. Colorado (USA). This development was supported by national funding agencies: CSA (Canada); NAOC (China); CEA, CNES, CNRS (France); ASI (Italy); MCINN (Spain); SNSB (Sweden); STFC, UKSA (UK); and NASA (USA).  This research was supported by the Natural Sciences and Engineering Research Council of Canada.

\bibliography{biblio.bib}


\appendix

\section{Source Catalogue} 
\label{Catalogue}

Tables~\ref{table:0303}--\ref{table:2349} comprise the list of LABOCA-detected sources presented in this paper, with spectroscopic (when available) or photometric redshifts quoted. We group the sources by field and then separate into four parts: those that fall within our radial cutoff, those that are within our radial cutoff and deboosted to the lower limit of the background prior, those that fall outside of radial cutoff, and those that are found with a lower SNR cut of 3$\sigma$.

\begin{table*}
\caption{SPT0303$-$59; $z_{\text{phot}}$\,=\,$3.33 \pm 0.37$.}
\begin{threeparttable}
\begin{tabular}{lllrrrrr}
    \hline
    Source name & Right Ascension & Declination & \centered{$S_{870}$} &  \centered{$S_{870}^{\rm deb}$} & \centered{$S_{500}$} & \centered{$S_{350}$} & \centered{$S_{250}$}\Tstrut\\
    & [J2000] & [J2000] & \centered{[mJy]} & \centered{[mJy]} & \centered{[mJy]} & \centered{[mJy]} & \centered{[mJy]}\Bstrut\\
    \hline
    \multicolumn{8}{c}{Sources within 240\,arcsec}\\
    \hline
    \Tstrut
    \input{0303/latexG}
    \hline
\end{tabular}
\begin{tablenotes}
\item[deb] Deboosted flux densities, 98\,per cent confidence upper limits are quoted when deboosted to the lower limits of the prior.
\item[1] Central core where \textit{Herschel}-SPIRE data are confused.
\end{tablenotes}
\end{threeparttable}
\label{table:0303}
\end{table*}

\begin{table*}
\caption{SPT0311$-$58; $z_{\text{spec}}\,{=}\,6.9011$.}
\begin{threeparttable}
\begin{tabular}{lllrrrrr}
    \hline
    Source name & Right Ascension & Declination & \centered{$S_{870}$} &  \centered{$S_{870}^{\rm deb}$} & \centered{$S_{500}$} & \centered{$S_{350}$} & \centered{$S_{250}$}\Tstrut\\
    & [J2000] & [J2000] & \centered{[mJy]} & \centered{[mJy]} & \centered{[mJy]} & \centered{[mJy]} & \centered{[mJy]}\Bstrut\\
    \hline
    \multicolumn{8}{c}{Sources within 240\,arcsec}\\
    \hline
    \Tstrut
    \input{0311/latexG}
    \hline
\end{tabular}
\begin{tablenotes}
\item[deb] Deboosted flux densities, 98\,per cent confidence upper limits are quoted when deboosted to the lower limits of the prior.
\item[1] Central core where \textit{Herschel}-SPIRE data are confused.
\end{tablenotes}
\end{threeparttable}
\label{table:0311}
\end{table*}

\begin{table*}
\caption{SPT0348$-$62; $z_{\text{spec}}\,{=}\,5.6541$.}
\begin{threeparttable}
\begin{tabular}{lllrrrrr}
    \hline
    Source name & Right Ascension & Declination & \centered{$S_{870}$} &  \centered{$S_{870}^{\rm deb}$} & \centered{$S_{500}$} & \centered{$S_{350}$} & \centered{$S_{250}$}\Tstrut\\
    & [J2000] & [J2000] & \centered{[mJy]} & \centered{[mJy]} & \centered{[mJy]} & \centered{[mJy]} &    
    \centered{[mJy]}\Bstrut\\
    \hline
    \multicolumn{8}{c}{Sources within 240\,arcsec}\\
    \hline
    \Tstrut
    \input{0348/latexG}
    \hline
\end{tabular}
\begin{tablenotes}
\item[deb] Deboosted flux densities, 98\,per cent confidence upper limits are quoted when deboosted to the lower limits of the prior.
\item[1] Central core where \textit{Herschel}-SPIRE data are confused.
\end{tablenotes}
\end{threeparttable}
\label{table:0348}
\end{table*}

\begin{table*}
\caption{SPT0457$-$49; $z_{\text{spec}}\,{=}\,3.9875$.}
\begin{threeparttable}
\begin{tabular}{lllrrrrr}
    \hline
    Source name & Right Ascension & Declination & \centered{$S_{870}$} &  \centered{$S_{870}^{\rm deb}$} & \centered{$S_{500}$} & \centered{$S_{350}$} & \centered{$S_{250}$}\Tstrut\\
    & [J2000] & [J2000] & \centered{[mJy]} & \centered{[mJy]} & \centered{[mJy]} & \centered{[mJy]} & \centered{[mJy]}\Bstrut\\
    \hline
    \multicolumn{8}{c}{Sources within 240\,arcsec}\\
    \hline
    \Tstrut
    \input{0457/latexG}
    \hline
\end{tabular}
\begin{tablenotes}
\item[deb] Deboosted flux densities, 98\,per cent confidence upper limits are quoted when deboosted to the lower limits of the prior.
\item[1] Central core where \textit{Herschel}-SPIRE data are confused.
\end{tablenotes}
\end{threeparttable}
\label{table:0457}
\end{table*}

\begin{table*}
\caption{SPT0553$-$50; $z_{\text{spec}}\,{=}\,5.323$.}
\begin{threeparttable}
\begin{tabular}{lllrrrrr}
    \hline
    Source name & Right Ascension & Declination & \centered{$S_{870}$} &  \centered{$S_{870}^{\rm deb}$} & \centered{$S_{500}$} & \centered{$S_{350}$} & \centered{$S_{250}$}\Tstrut\\
    & [J2000] & [J2000] & \centered{[mJy]} & \centered{[mJy]} & \centered{[mJy]} & \centered{[mJy]} & \centered{[mJy]}\Bstrut\\
    \hline
    \multicolumn{8}{c}{Sources within 240\,arcsec}\\
    \hline
    \Tstrut
    \input{0553/latexG}
    \hline
\end{tabular}
\begin{tablenotes}
\item[deb] Deboosted flux densities, 98\,per cent confidence upper limits are quoted when deboosted to the lower limits of the prior.
\item[1] Central core where \textit{Herschel}-SPIRE data are confused.
\end{tablenotes}
\end{threeparttable}
\label{table:0553}
\end{table*}

\begin{table*}
\caption{SPT2018$-$45; $z_{\text{phot}}\,{=}\,3.18\pm 0.39$.}
\begin{threeparttable}
\begin{tabular}{lllrrrrr}
    \hline
    Source name & Right Ascension & Declination & \centered{$S_{870}$} &  \centered{$S_{870}^{\rm deb}$} & \centered{$S_{500}$} & \centered{$S_{350}$} & \centered{$S_{250}$}\Tstrut\\
    & [J2000] & [J2000] & \centered{[mJy]} & \centered{[mJy]} & \centered{[mJy]} & \centered{[mJy]} & \centered{[mJy]}\Bstrut\\
    \hline
    \multicolumn{8}{c}{Sources within 240\,arcsec}\\
    \hline
    \Tstrut
    \input{2018/latexG}
    \hline
\end{tabular}
\begin{tablenotes}
\item[deb] Deboosted flux densities, 98\,per cent confidence upper limits are quoted when deboosted to the lower limits of the prior.
\item[1] Central core where \textit{Herschel}-SPIRE data are confused.
\end{tablenotes}
\end{threeparttable}
\label{table:2018}
\end{table*}

\begin{table*}
\caption{SPT2052$-$56; $z_{\text{spec}}\,{=}\,4.257$.}
\begin{threeparttable}
\begin{tabular}{lllrrrrr}
    \hline
    Source name & Right Ascension & Declination & \centered{$S_{870}$} &  \centered{$S_{870}^{\rm deb}$} & \centered{$S_{500}$} & \centered{$S_{350}$} & \centered{$S_{250}$}\Tstrut\\
    & [J2000] & [J2000] & \centered{[mJy]} & \centered{[mJy]} & \centered{[mJy]} & \centered{[mJy]} & \centered{[mJy]}\Bstrut\\
    \hline
    \multicolumn{8}{c}{Sources within 240\,arcsec}\\
    \hline
    \Tstrut
    \input{2052/latexG}
    \hline
\end{tabular}
\begin{tablenotes}
\item[deb] Deboosted flux densities, 98\,per cent confidence upper limits are quoted when deboosted to the lower limits of the prior.
\item[1] Central core where \textit{Herschel}-SPIRE data are confused.
\end{tablenotes}
\end{threeparttable}
\label{table:2052}
\end{table*}

\begin{table*}
\caption{SPT2335$-$53; $z_{\text{spec}}\,{=}\,4.7555$.}
\begin{threeparttable}
\begin{tabular}{lllrrrrr}
    \hline
    Source name & Right Ascension & Declination & \centered{$S_{870}$} &  \centered{$S_{870}^{\rm deb}$} & \centered{$S_{500}$} & \centered{$S_{350}$} & \centered{$S_{250}$}\Tstrut\\
    & [J2000] & [J2000] & \centered{[mJy]} & \centered{[mJy]} & \centered{[mJy]} & \centered{[mJy]} & \centered{[mJy]}\Bstrut\\
    \hline
    \multicolumn{8}{c}{Sources within 240\,arcsec}\\
    \hline
    \Tstrut
    \input{2335/latexG}
    \hline
\end{tabular}
\begin{tablenotes}
\item[deb] Deboosted flux densities, 98\,per cent confidence upper limits are quoted when deboosted to the lower limits of the prior.
\item[1] Central core where \textit{Herschel}-SPIRE data are confused.
\end{tablenotes}
\end{threeparttable}
\label{table:2335}
\end{table*}

\begin{table*}
\caption{SPT2349$-$56; $z_{\text{spec}}\,{=}\,4.3020$.}
\begin{threeparttable}
\begin{tabular}{lllrrrrr}
    \hline
    Source name & Right Ascension & Declination & \centered{$S_{870}$} &  \centered{$S_{870}^{\rm deb}$} & \centered{$S_{500}$} & \centered{$S_{350}$} & \centered{$S_{250}$}\Tstrut\\
    & [J2000] & [J2000] & \centered{[mJy]} & \centered{[mJy]} & \centered{[mJy]} & \centered{[mJy]} & \centered{[mJy]}\Bstrut\\
    \hline
    \multicolumn{8}{c}{Sources within 240\,arcsec}\\
    \hline
    \Tstrut
    \input{2349/latexG}
    \hline
\end{tabular}
\begin{tablenotes}
\item[deb] Deboosted flux densities, 98\,per cent confidence upper limits are quoted when deboosted to the lower limits of the prior.
\item[1] Southern core where \textit{Herschel}-SPIRE data are confused.
\item[2] Northern core where \textit{Herschel}-SPIRE data are confused.
\end{tablenotes}
\end{threeparttable}
\label{table:2349}
\end{table*}

\newpage

\section{LABOCA Maps and \textit{Herschel} Flux Densities}
\label{Maps}

\begin{figure*}
    \includegraphics[width = \width{}]{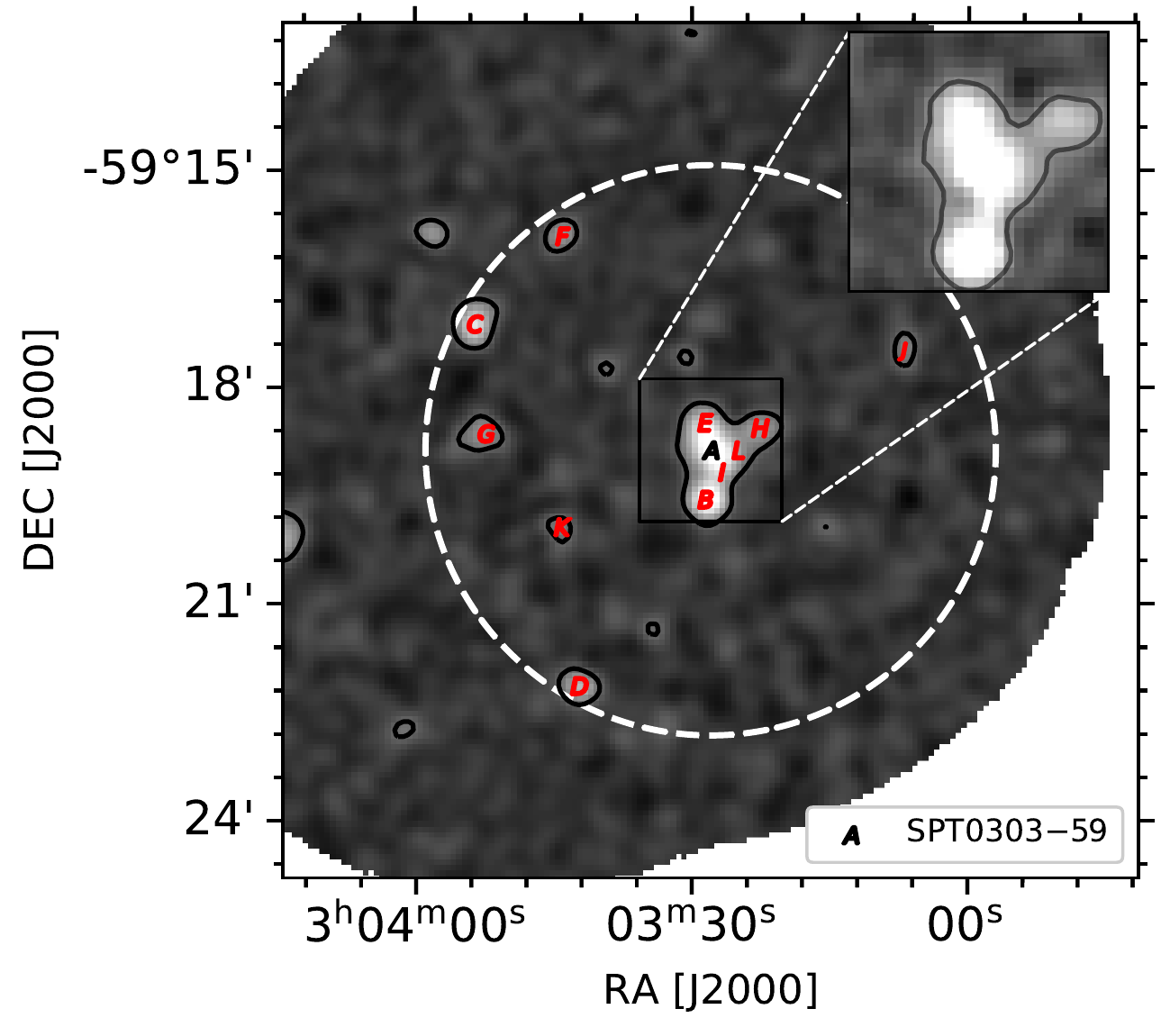}
    \includegraphics[width = \width{}]{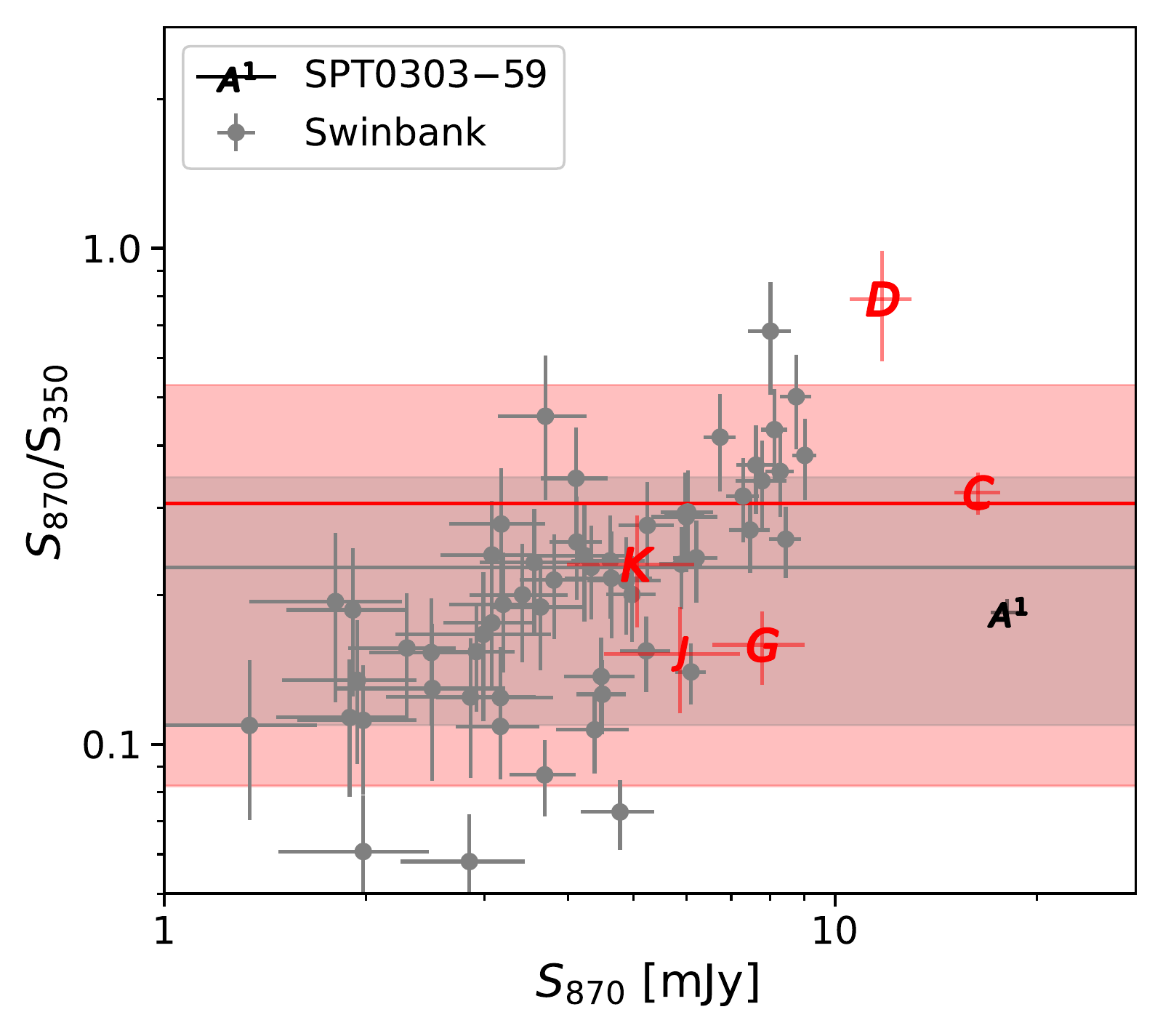}
    \includegraphics[width = \width{}]{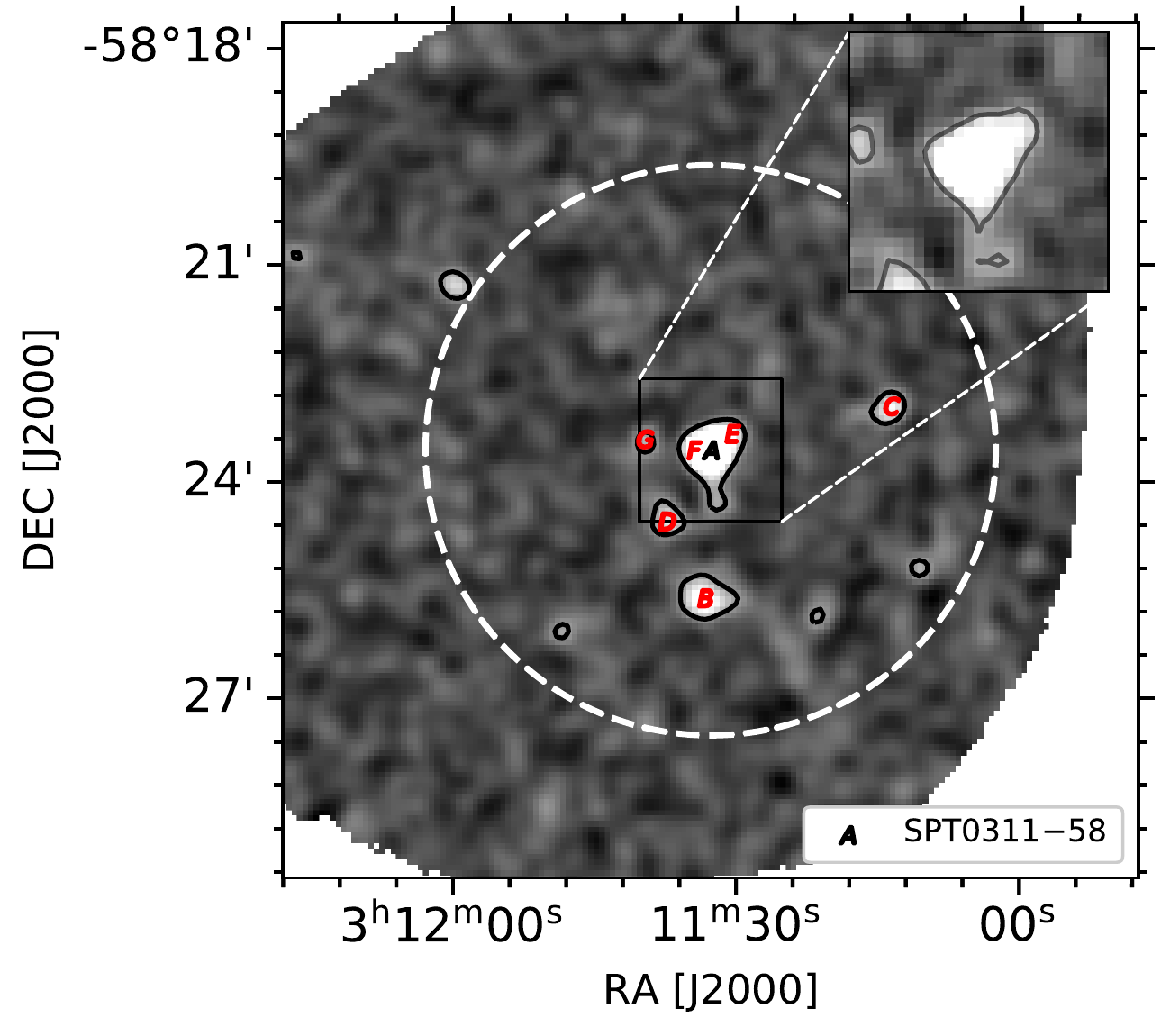}
    \includegraphics[width = \width{}]{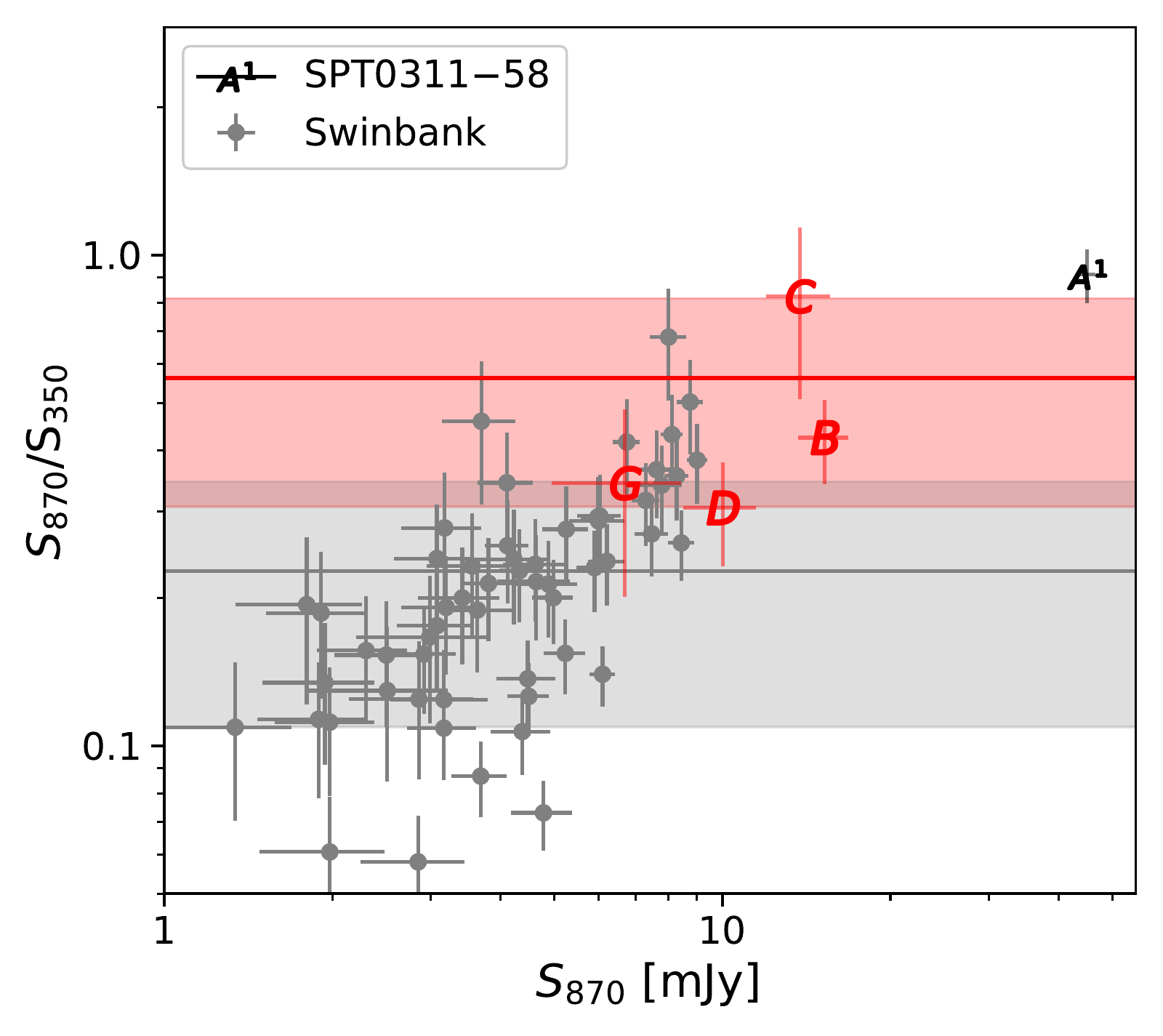}
    \includegraphics[width = \width{}]{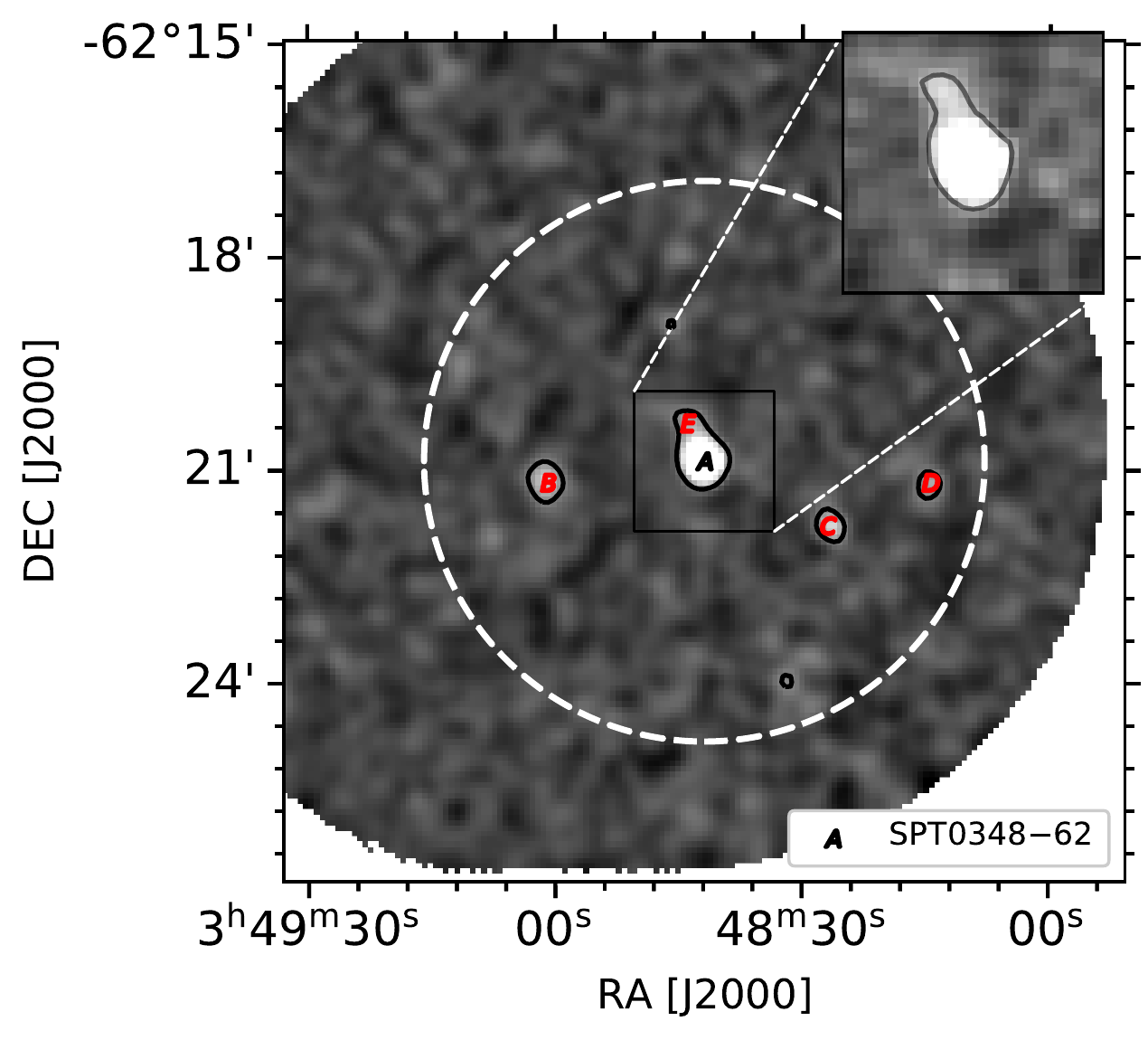}
    \includegraphics[width = \width{}]{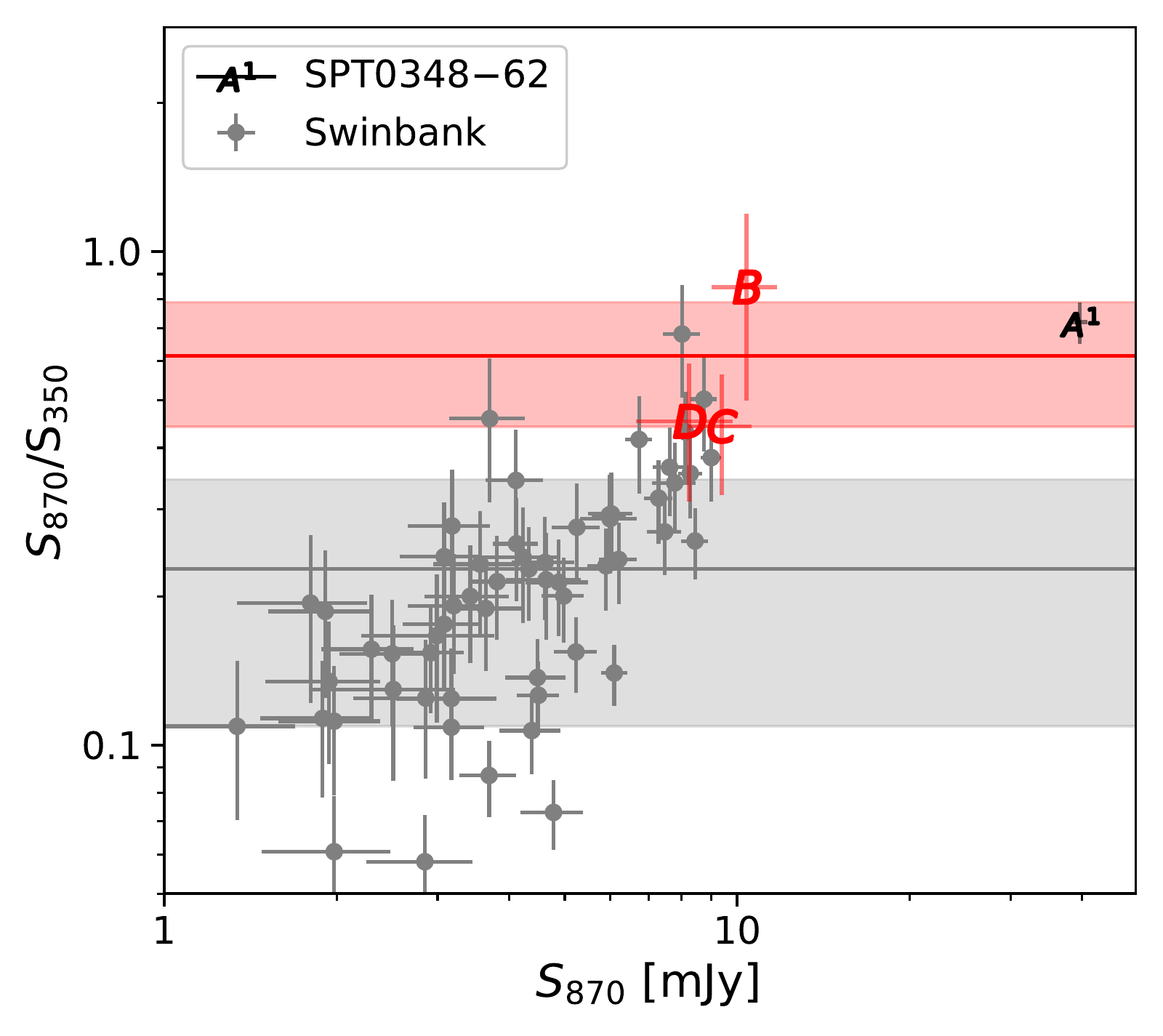}
    \caption{Spatial distributions and colours of sources in each field. \textit{Left}: The greyscale image shows the 18.6-arcsec-convolved 870-$\mu$m LABOCA map. The objects overlaid on these images are the sources within our 240\,arcsec cutoff labelled accordingly to Appendix~\ref{Catalogue}. Contours corresponding to 3.7$\sigma$ are displayed in black, while the white dotted-circle represents the radial cutoff. Insets show enlarged areas of the central core (12\,arcsec on a side) convolved LABOCA map. \textit{Right}: Flux density ratio versus flux density plot. The sources are labelled the same as on the LABOCA maps, but we show the integrated colour-brightness point for the extended central core with integer superscripts. We compare the average colours of our fields to the sample of SMGs from \citet{Swinbank2014} (light grey) and show that our sources have higher colours.}
    \label{fig: 0303}
\end{figure*}

Figure~\ref{fig: 0303} shows the LABOCA beam convolved (18.6\,arcsec Gaussian beam) SNR maps (left) and colour-flux plots (right) for each of the nine fields. We show contours at our detection threshold of 3.7$\sigma$. The sources are labelled accordingly to Appendix~\ref{Catalogue}, where only the sources within the 240\,arcsec (1.3--1.9\,Mpc) cutoff are shown, and the central sources labelled with black A's. The white circles depict the radial cutoff of 240\,arcsec used in our selection criteria. A small cutout of each field's central core is shown, with the image and contour corresponding to a less smoothed version of the LABOCA SNR maps (12\,arcsec Gaussian beam) highlighting the interesting substructure seen in our data. For the colour-brightness, the $S_{870}/S_{350}$ versus $S_{870}$ flux densities are shown with the same labels as the SNR maps with background SMGs from \citet{Swinbank2014} plotted in light grey. The black horizontal lines and the shaded regions show the background SMGs' averages and standard deviations, respectively, while the red horizontal lines and shaded areas show the averages and standard deviations of our sources. For each field, we show the integrated colour and brightness of the extended central core(s) as labelled in Appendix~\ref{Catalogue} (identified by integer superscripts) due to the extended emission in the \textit{Herschel}-SPIRE bands. There are some sources with no measured 350-$\mu$m emissions (or non-positive), so we exclude these sources from the colour-brightness plots.

\renewcommand{\thefigure}{B\arabic{figure} (Cont.)}
\addtocounter{figure}{-1}
\begin{figure*}
    \includegraphics[width = \width{}]{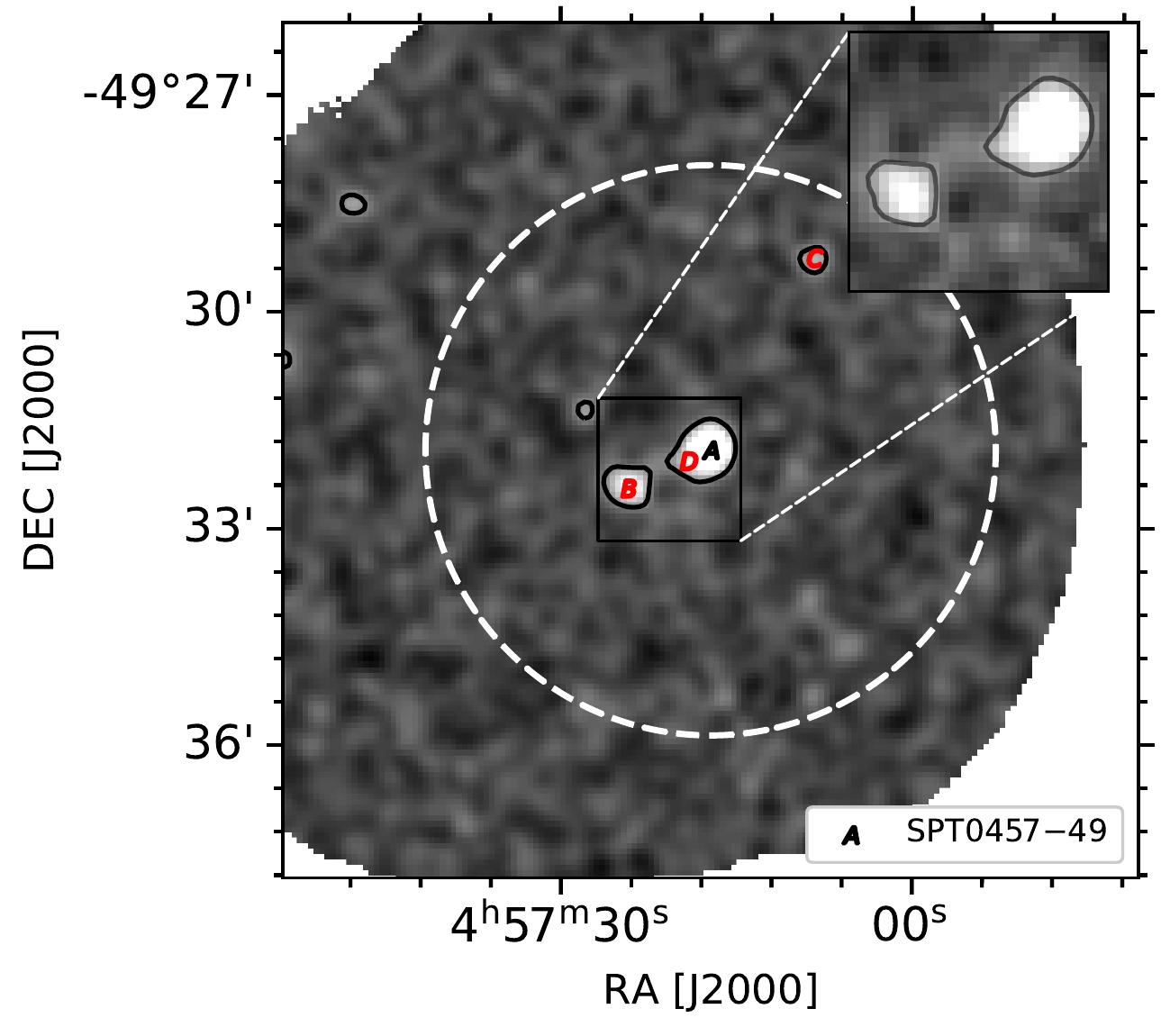}
    \includegraphics[width = \width{}]{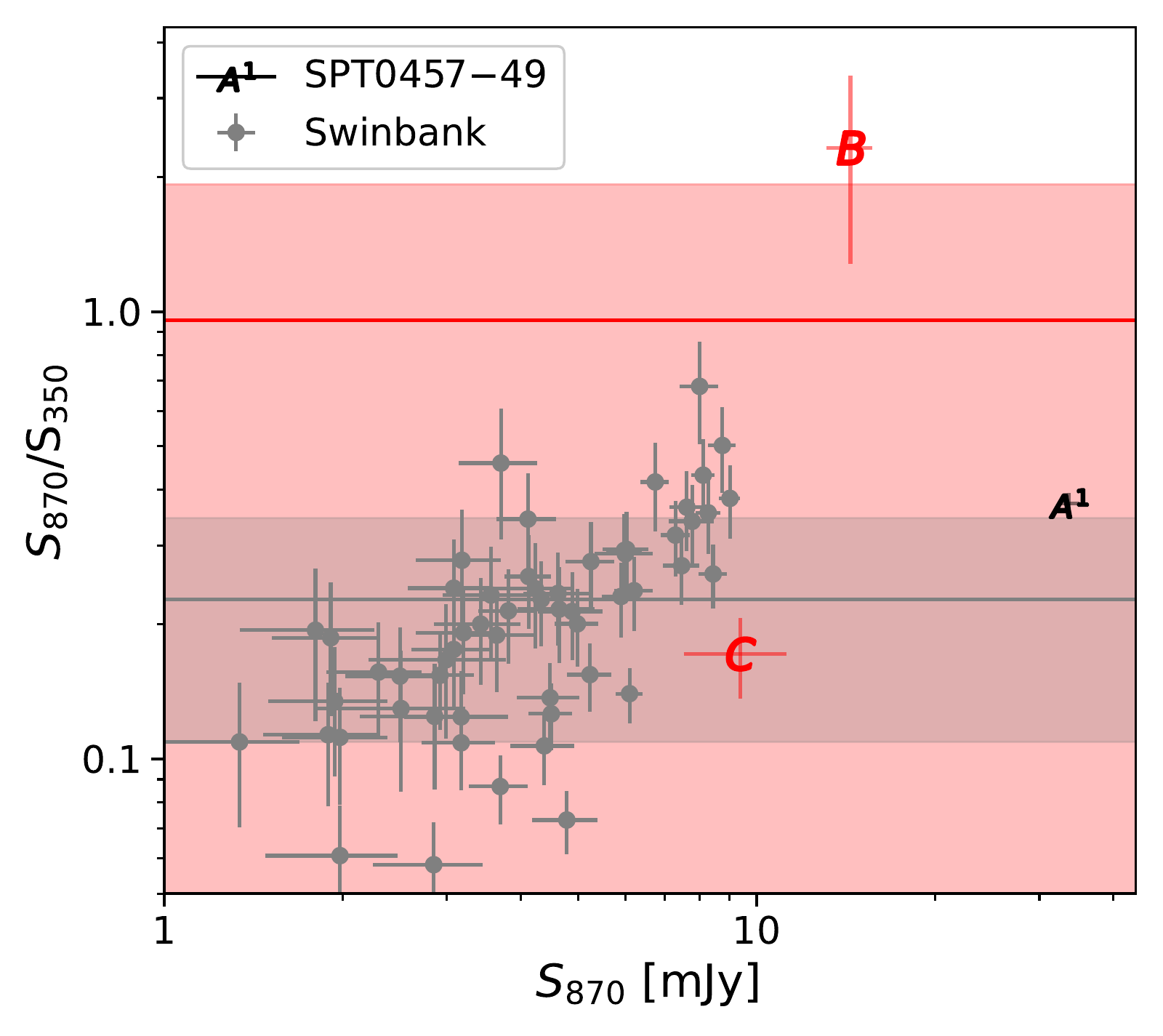}
    \includegraphics[width = \width{}]{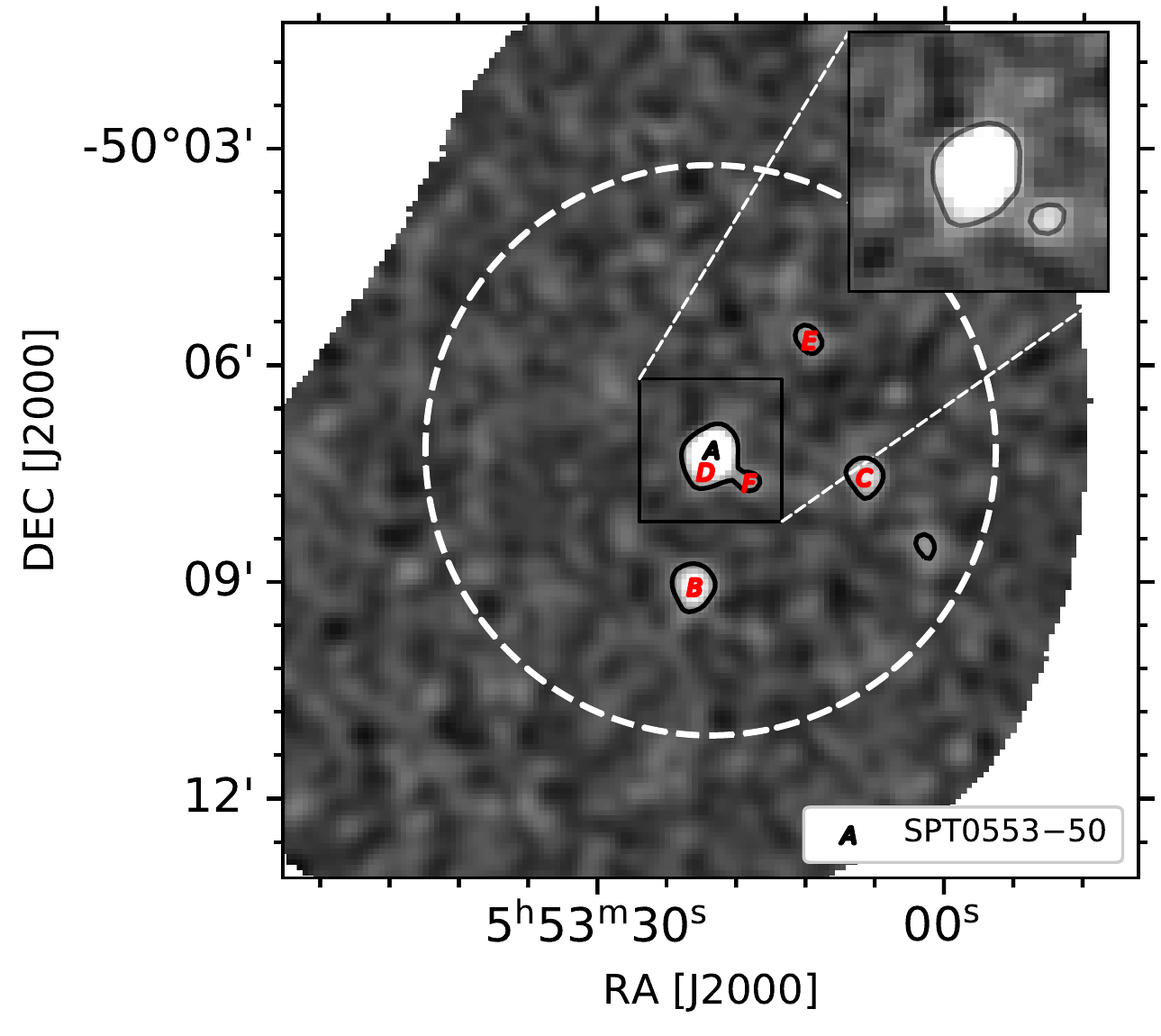}
    \includegraphics[width = \width{}]{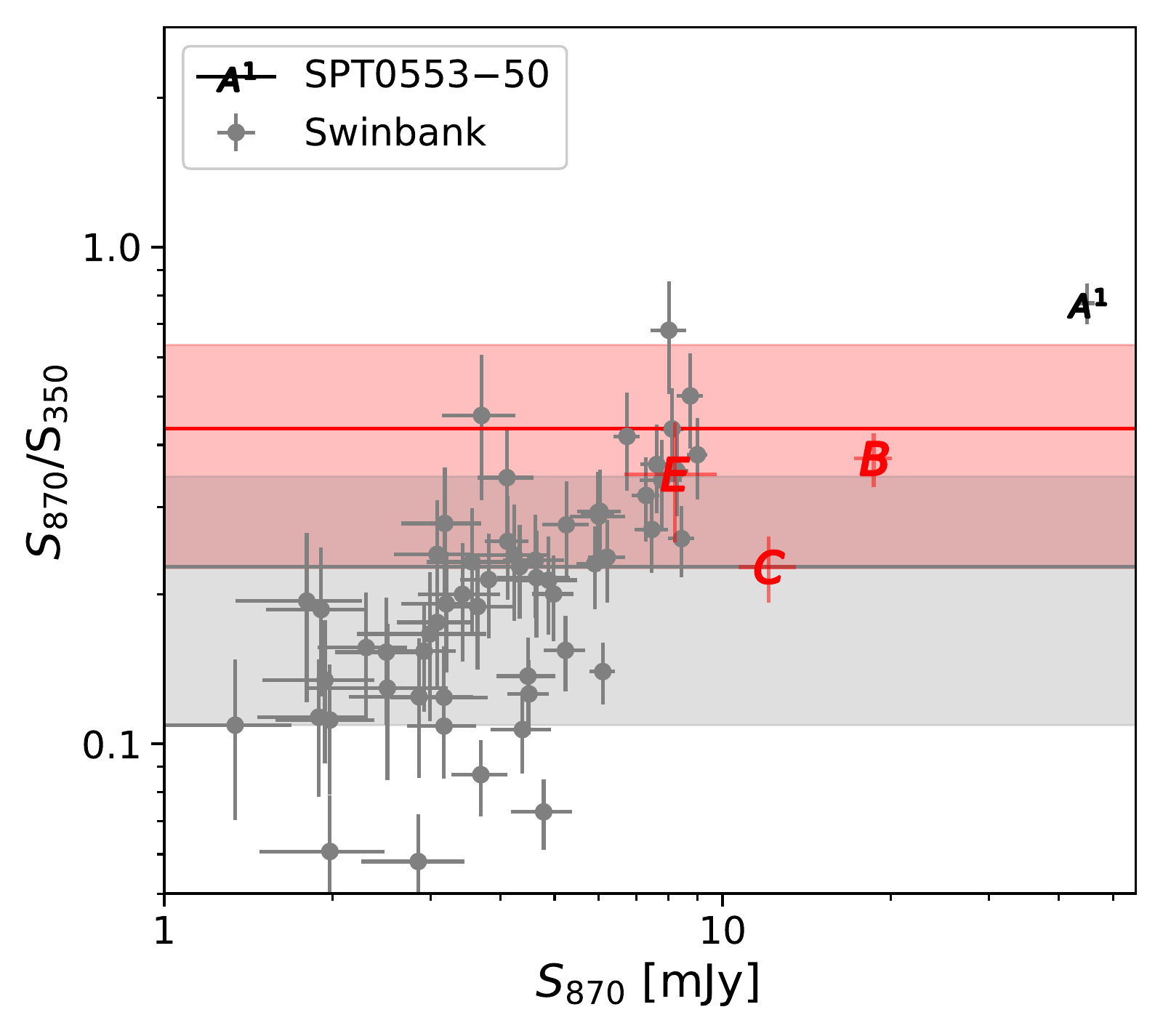}
    \includegraphics[width = \width{}]{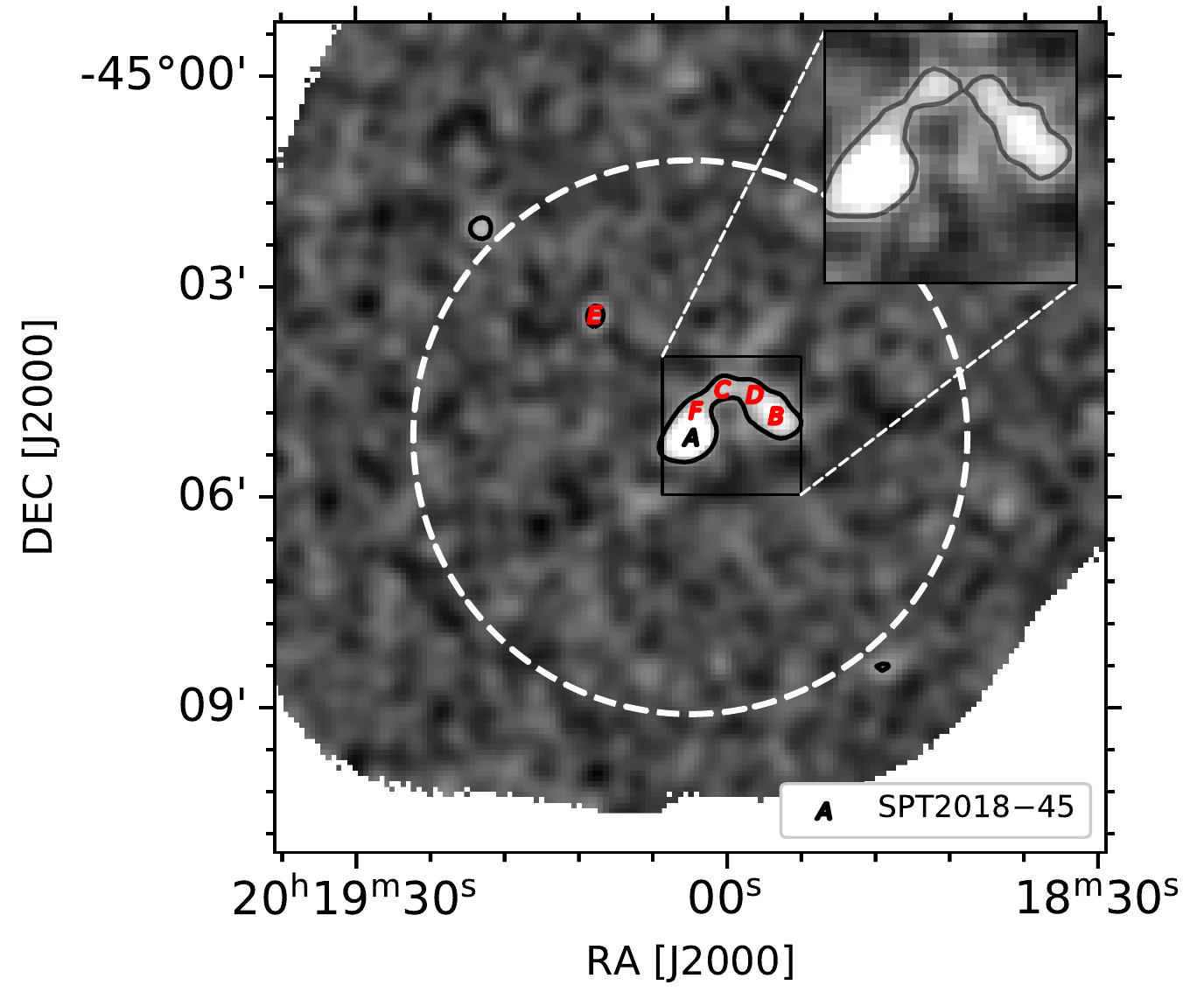}
    \includegraphics[width = \width{}]{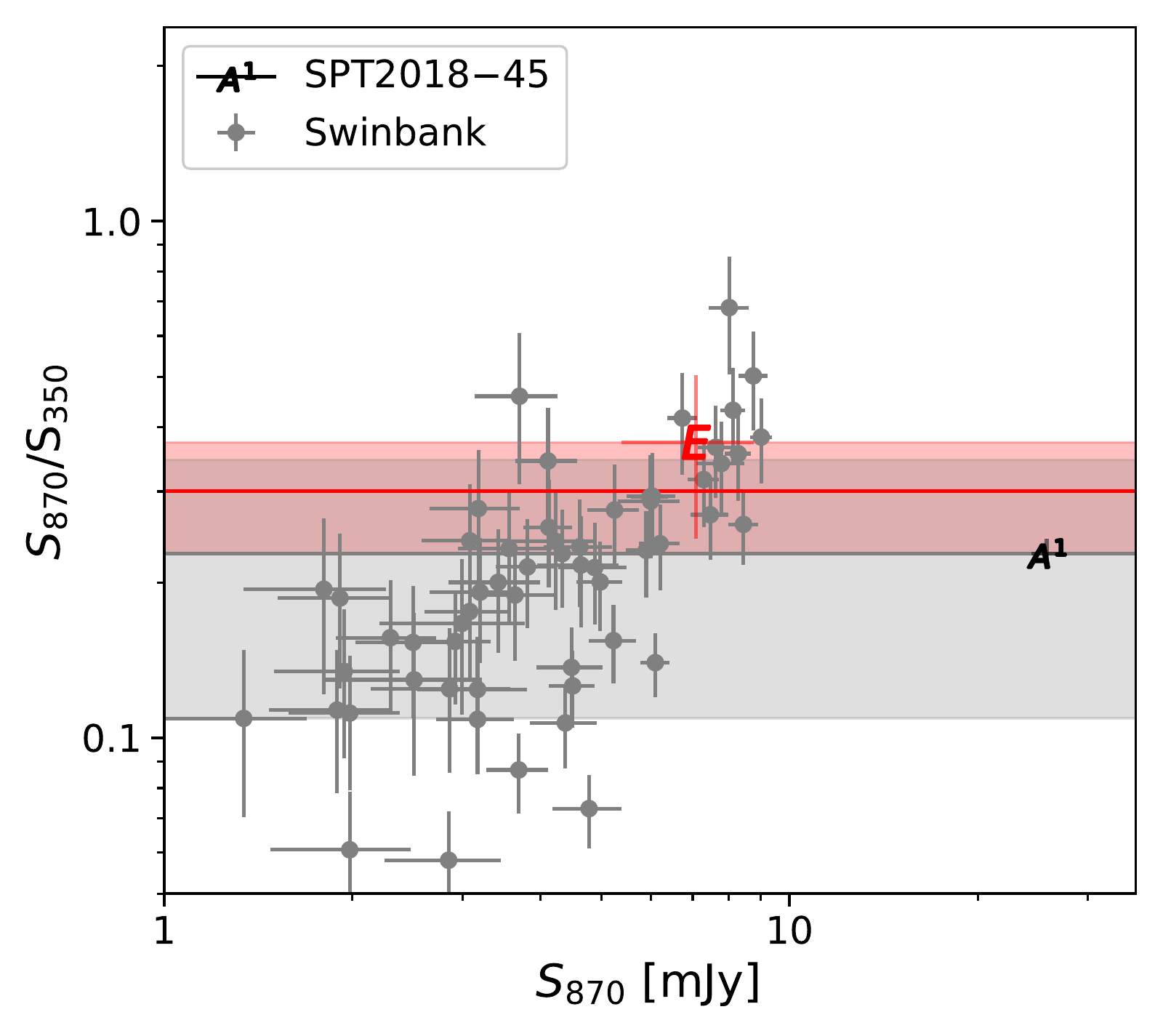}
    \caption{}
\end{figure*}
\renewcommand{\thefigure}{\arabic{figure}}

\renewcommand{\thefigure}{B\arabic{figure} (Cont.)}
\addtocounter{figure}{-1}
\begin{figure*}
    \includegraphics[width = \width{}]{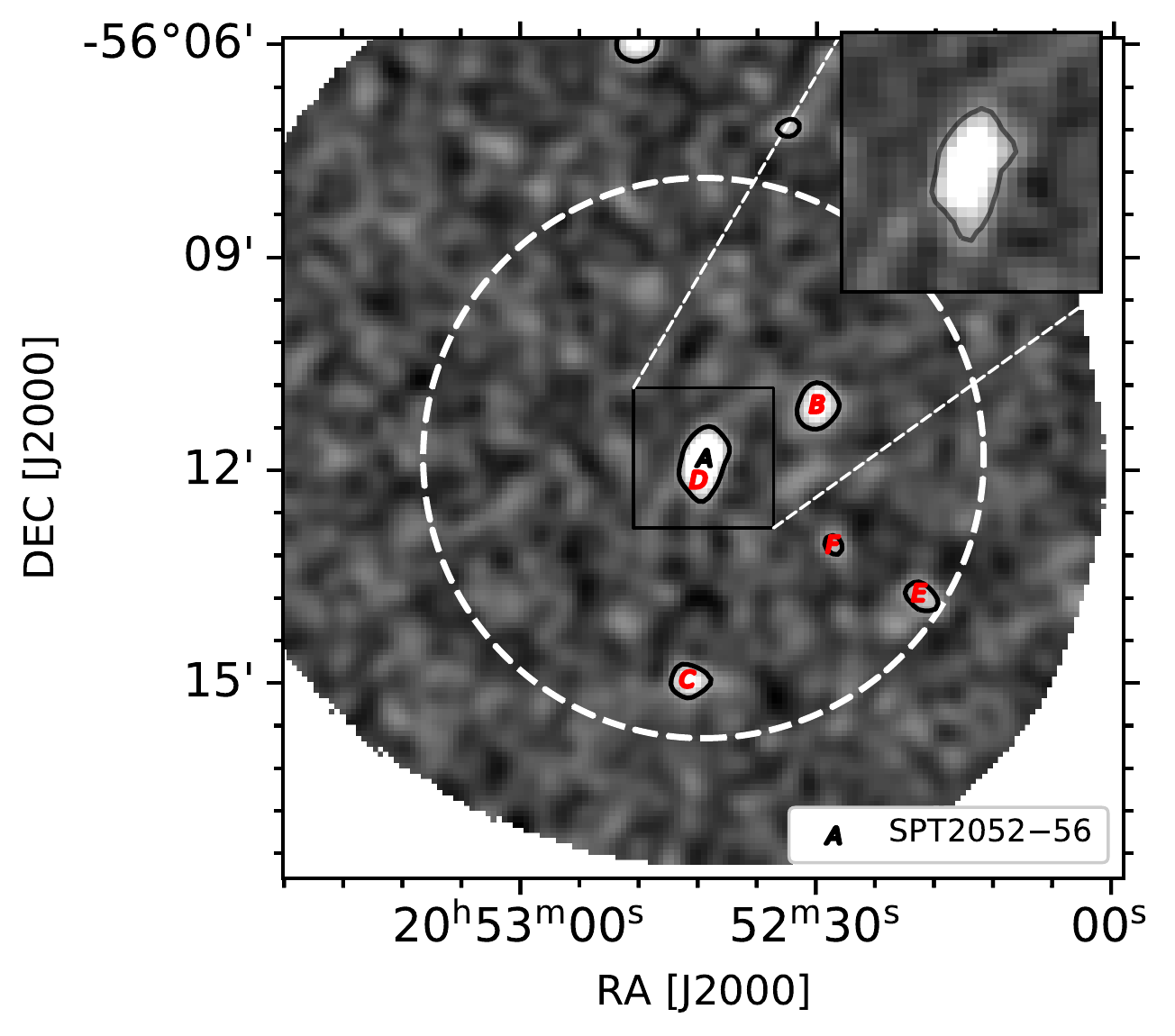}
    \includegraphics[width = \width{}]{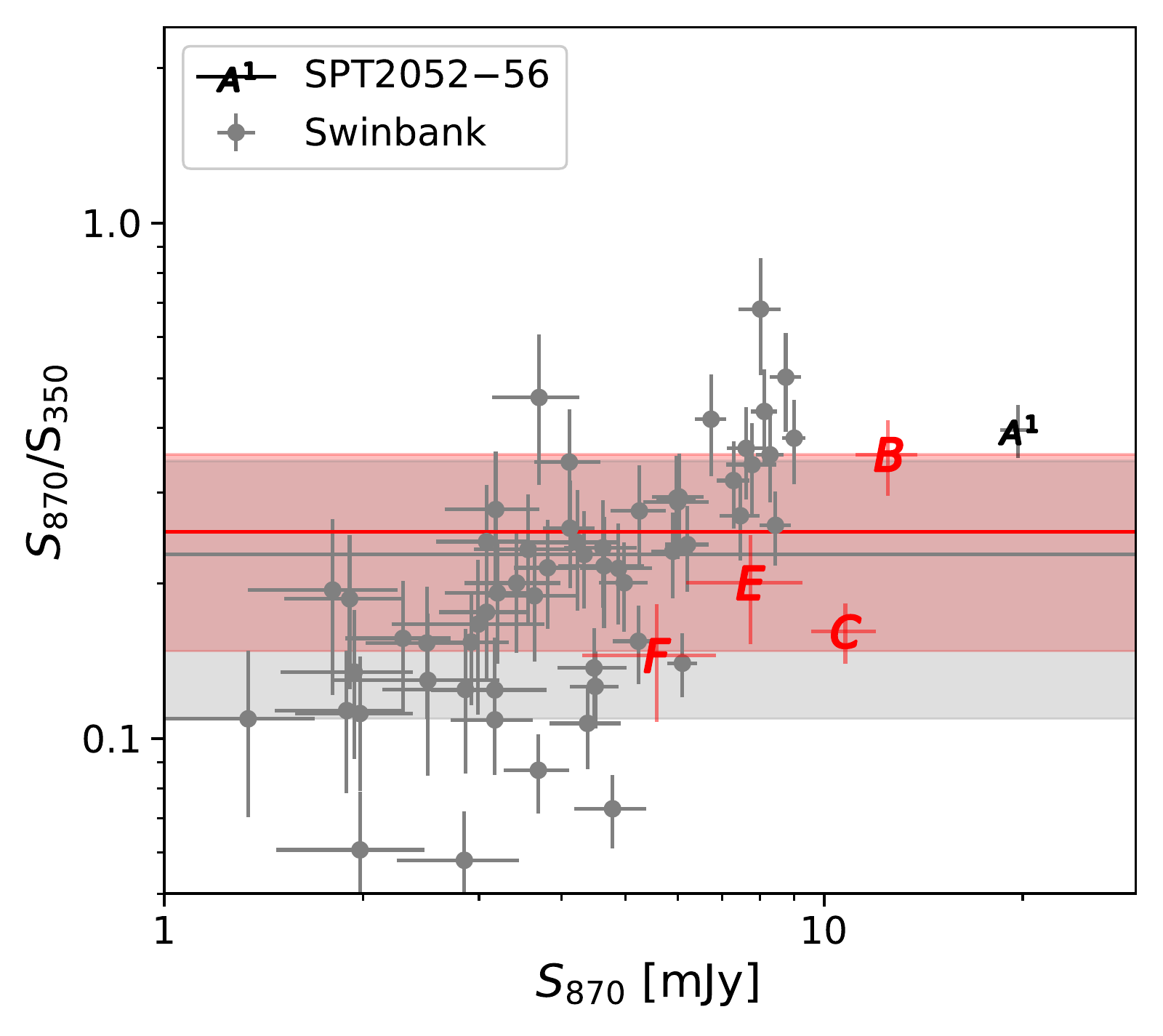}
    \includegraphics[width = \width{}]{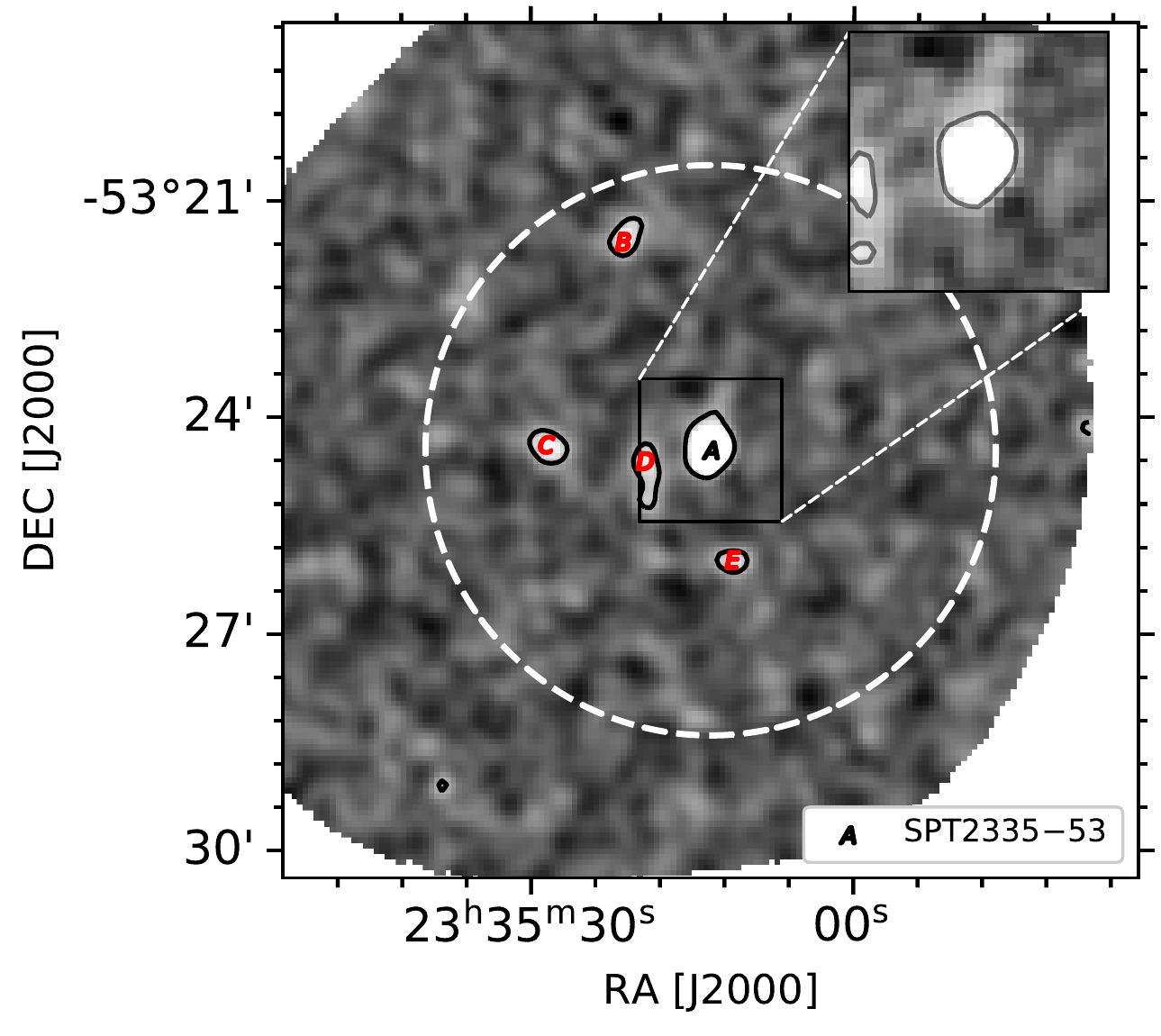}
    \includegraphics[width = \width{}]{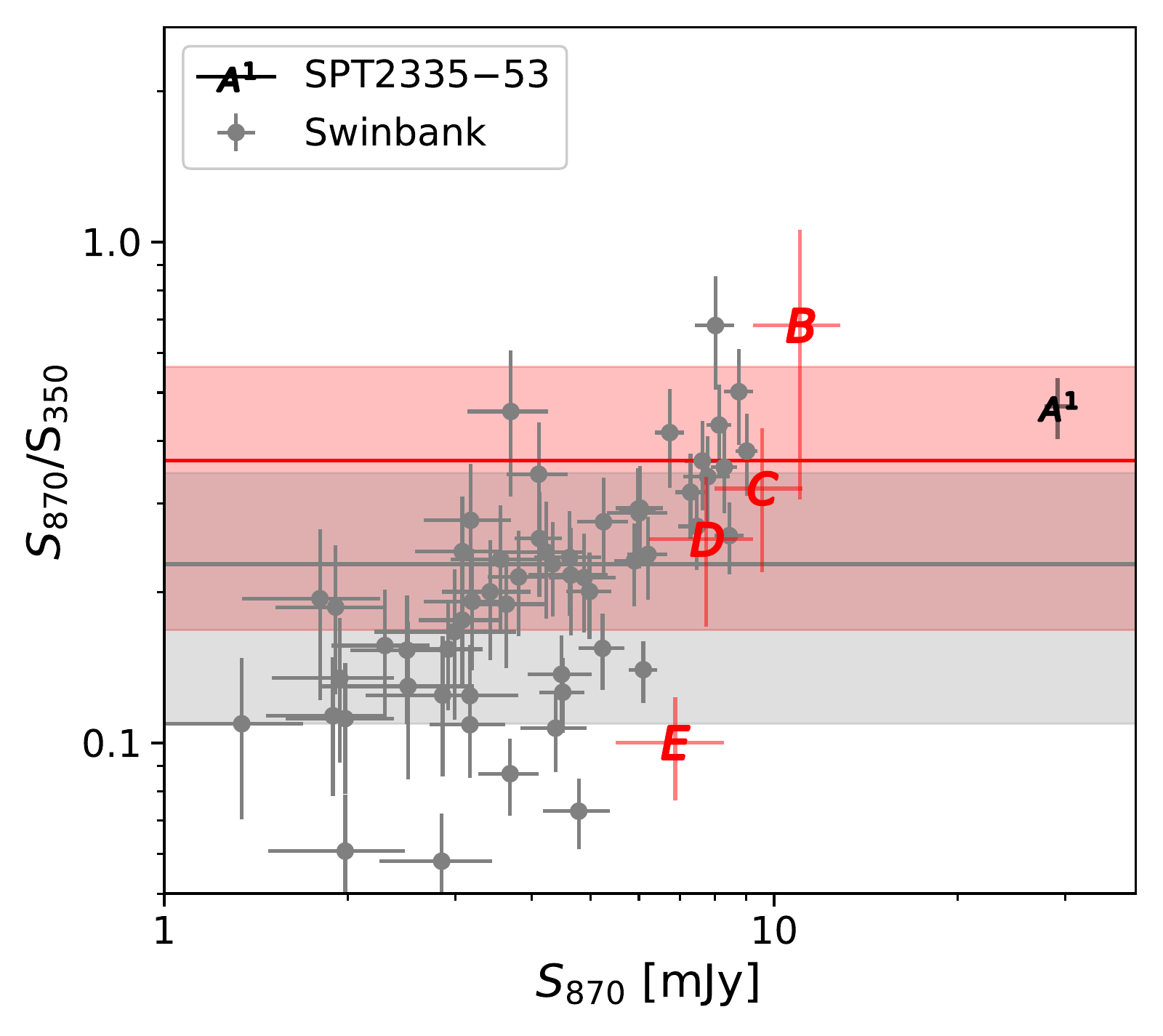}
    \includegraphics[width = \width{}]{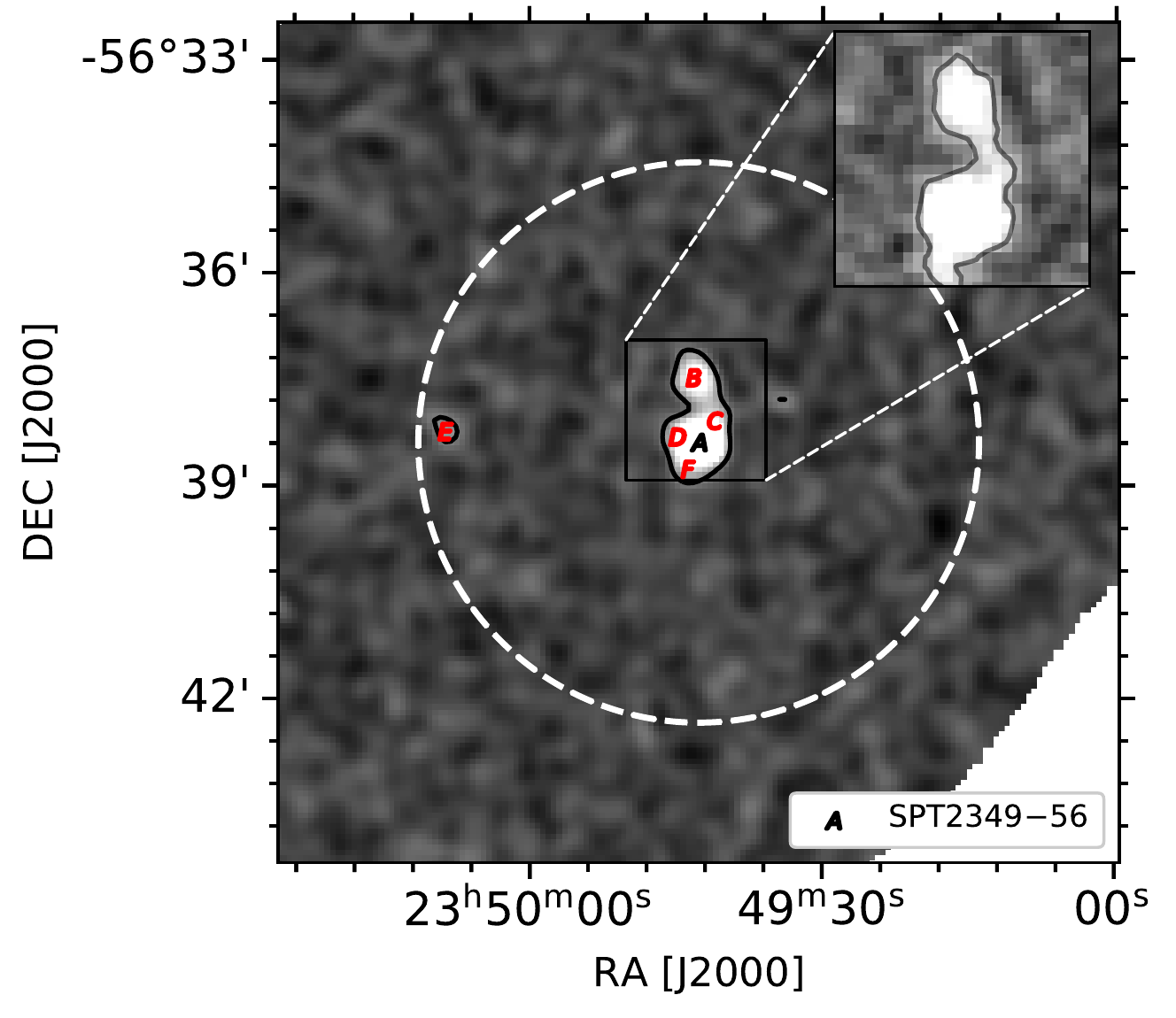}
    \includegraphics[width = \width{}]{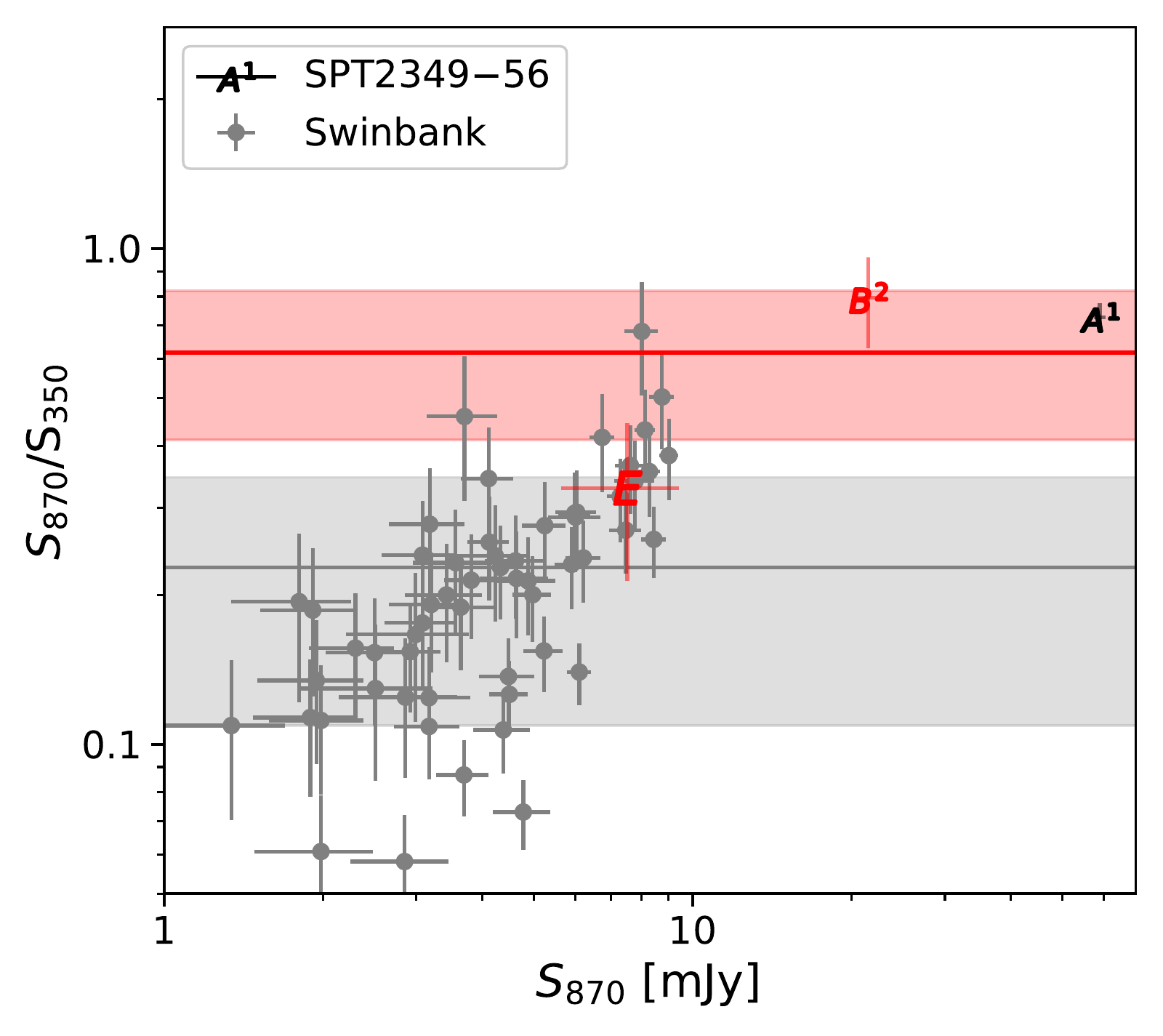}
    \caption{}
\end{figure*}
\renewcommand{\thefigure}{\arabic{figure}}


\bsp    
\label{lastpage}
\end{document}

%% file: instrument
SPT0303$-$59 & 03:03:28 & $-$59:18:52 & 31 & 0.99 & 145\\
SPT0311$-$58 & 03:11:33 & $-$58:23:34 & 15 & 1.46 & 146\\
SPT0348$-$62 & 03:48:42 & $-$62:20:52 & 13 & 1.21 & 138\\
SPT0457$-$49 & 04:57:17 & $-$49:31:55 & 17 & 1.36 & 144\\
SPT0553$-$50 & 05:53:20 & $-$50:07:11 & 23 & 1.41 & 133\\
SPT2018$-$45 & 20:19:03 & $-$45:05:09 & 15 & 1.39 & 153\\
SPT2052$-$56 & 20:52:41 & $-$56:11:50 & 31 & 1.15 & 138\\
SPT2335$-$53 & 23:35:13 & $-$53:24:27 & 19 & 1.38 & 140\\
SPT2349$-$56 & 23:49:43 & $-$56:38:24 & 20 & 1.40 & 225\\

%% file: SFRlatex
SPT0303$-$59 & 03:03:28 & $-$59:18:52 & 3.3$^5$ & 20.5$\pm$4.0 & 65.19$\pm$2.47 & 15.7 & 2.4 & 6.1\\
SPT0311$-$58 & 03:11:33 & $-$58:23:34 & 6.9 & 17.8$\pm$4.1 & 61.11$\pm$10.21 & 10.9 & 4.6 & 12.0\\
SPT0348$-$62 & 03:48:42 & $-$62:20:52 & 5.7 & 17.1$\pm$5.8 & 47.28$\pm$8.23 & 7.8 & 4.1 & 6.1\\
SPT0457$-$49 & 04:57:17 & $-$49:31:55 & 4.0 & 6.8$\pm$4.4 & 42.0$\pm$1.96 & 7.8 & 4.0 & 3.7\\
SPT0553$-$50 & 05:53:20 & $-$50:07:11 & 5.3 & 9.6$\pm$6.8 & 60.61$\pm$2.58 & 10.5 & 4.7 & 7.5\\
SPT2018$-$45 & 20:19:03 & $-$45:05:09 & 3.2$^5$ & 22.1$\pm$4.6 & 61.56$\pm$10.0 & 9.2 & 3.4 & 3.4\\
SPT2052$-$56 & 20:52:41 & $-$56:11:50 & 4.3 & 16.0$\pm$3.4 & 28.01$\pm$7.81 & 7.4 & 2.3 & 3.9\\
SPT2335$-$53 & 23:35:13 & $-$53:24:27 & 4.8 & 11.6$\pm$2.7 & 29.13$\pm$1.41 & 7.0 & 3.2 & 4.3\\
SPT2349$-$56 & 23:49:43 & $-$56:38:24 & 4.3 & 19.0$\pm$3.0 & 105.84$\pm$8.22 & 12.9 & 6.7 & 6.8\\

%% file: 0303/latexG
A$^{1}$ & 03:03:27.97 & $-$59:18:52.39 & 18.3 $\pm$ 1.0 & $18.1^{+1.0}_{-1.0}$ & 87.2 $\pm$ 4.4 & 78.2 $\pm$ 4.1 & 44.0 $\pm$ 4.3\\
B$^{1}$ & 03:03:28.56 & $-$59:19:33.39 & 17.2 $\pm$ 0.9 & $17.0^{+0.9}_{-0.9}$ & $--$ & $--$ & $--$\\
C & 03:03:53.54 & $-$59:17:07.61 & 16.8 $\pm$ 1.3 & $16.3^{+1.3}_{-1.3}$ & 50.0 $\pm$ 4.1 & 50.8 $\pm$ 3.0 & 39.7 $\pm$ 4.1\\
D & 03:03:42.26 & $-$59:22:08.28 & 12.2 $\pm$ 1.2 & $11.7^{+1.2}_{-1.2}$ & 17.1 $\pm$ 4.3 & 14.9 $\pm$ 3.4 & 17.4 $\pm$ 3.8\\
E$^{1}$ & 03:03:28.56 & $-$59:18:29.61 & 11.9 $\pm$ 1.0 & $11.6^{+1.1}_{-1.0}$ & $--$ & $--$ & $--$\\
F & 03:03:44.02 & $-$59:15:54.72 & 9.1 $\pm$ 1.3 & $8.5^{+1.3}_{-1.3}$ & 3.4 $\pm$ 4.9 & $<$2.9 & $<$4.6\\
G & 03:03:52.36 & $-$59:18:38.72 & 8.4 $\pm$ 1.2 & $7.8^{+1.2}_{-1.2}$ & 44.6 $\pm$ 4.2 & 49.1 $\pm$ 3.0 & 55.6 $\pm$ 3.7\\
H$^{1}$ & 03:03:22.61 & $-$59:18:34.17 & 7.9 $\pm$ 1.0 & $7.5^{+1.0}_{-1.0}$ & $--$ & $--$ & $--$\\
I$^{1}$ & 03:03:26.77 & $-$59:19:10.61 & 7.4 $\pm$ 1.0 & $6.9^{+1.0}_{-1.0}$ & $--$ & $--$ & $--$\\
J & 03:03:07.16 & $-$59:17:30.39 & 6.7 $\pm$ 1.2 & $5.9^{+1.3}_{-1.4}$ & 35.3 $\pm$ 3.7 & 38.6 $\pm$ 3.0 & 33.7 $\pm$ 3.8\\
K & 03:03:44.04 & $-$59:19:56.17 & 5.7 $\pm$ 1.0 & $5.1^{+1.1}_{-1.1}$ & 23.2 $\pm$ 3.7 & 22.0 $\pm$ 3.0 & 16.0 $\pm$ 3.7\\
L$^{1}$ & 03:03:24.99 & $-$59:18:52.39 & 4.8 $\pm$ 1.0 & $4.1^{+1.1}_{-1.1}$ & $--$ & $--$ & $--$\\
\hline \multicolumn{8}{c}{Sources within 240\,arcsec and deboosted to zero}\\  \hline M & 03:03:30.94 & $-$59:17:34.94 & 4.8 $\pm$ 1.1 & $<$7.0 & 12.4 $\pm$ 3.3 & 20.4 $\pm$ 2.7 & 21.0 $\pm$ 3.5\\
N & 03:03:39.27 & $-$59:17:44.06 & 4.7 $\pm$ 1.1 & $<$6.8 & 11.6 $\pm$ 3.3 & 7.3 $\pm$ 2.8 & 0.5 $\pm$ 3.7\\
O & 03:03:34.51 & $-$59:21:22.72 & 4.3 $\pm$ 1.1 & $<$6.3 & 26.0 $\pm$ 3.6 & 42.4 $\pm$ 2.6 & 48.3 $\pm$ 3.5\\
P & 03:03:15.46 & $-$59:19:56.17 & 3.6 $\pm$ 1.0 & $<$5.3 & 14.5 $\pm$ 3.1 & 22.6 $\pm$ 2.7 & 16.9 $\pm$ 3.3\\
\hline \multicolumn{8}{c}{Sources outside 240\,arcsec}\\  \hline Q & 03:04:14.41 & $-$59:20:05.28 & 13.8 $\pm$ 1.6 & $13.0^{+1.6}_{-1.6}$ & 74.6 $\pm$ 4.7 & 79.1 $\pm$ 6.2 & 48.2 $\pm$ 6.3\\
R & 03:03:58.28 & $-$59:15:50.17 & 9.8 $\pm$ 1.4 & $9.0^{+1.4}_{-1.4}$ & 20.8 $\pm$ 4.4 & 15.8 $\pm$ 3.4 & 10.4 $\pm$ 4.2\\
S & 03:04:19.84 & $-$59:23:30.28 & 17.6 $\pm$ 3.8 & $<$23.7 & \dots & \dots & \dots\\
T & 03:03:00.09 & $-$59:13:33.50 & 16.6 $\pm$ 3.9 & $<$21.7 & 22.0 $\pm$ 4.7 & \dots & 6.6 $\pm$ 4.1\\
U & 03:04:16.19 & $-$59:19:47.06 & 7.9 $\pm$ 1.7 & $<$11.3 & 78.9 $\pm$ 5.0 & \dots & 22.7 $\pm$ 5.4\\
V & 03:03:30.36 & $-$59:13:06.16 & 7.6 $\pm$ 2.0 & $<$10.5 & 48.9 $\pm$ 9.1 & 48.4 $\pm$ 4.5 & 62.7 $\pm$ 6.7\\
W & 03:04:01.34 & $-$59:22:44.72 & 7.1 $\pm$ 1.6 & $<$10.2 & 37.1 $\pm$ 4.4 & 34.6 $\pm$ 4.2 & 27.3 $\pm$ 8.9\\
\hline \multicolumn{8}{c}{Sources detected at 3$\sigma$}\\  \hline X & 03:03:32.72 & $-$59:22:40.17 & 4.6 $\pm$ 1.3 & $<$6.4 & 23.3 $\pm$ 4.3 & 25.9 $\pm$ 3.3 & 19.8 $\pm$ 4.1\\
Y & 03:03:55.34 & $-$59:18:47.83 & 3.7 $\pm$ 1.2 & $<$5.2 & 31.1 $\pm$ 3.9 & 15.7 $\pm$ 2.9 & 14.8 $\pm$ 3.4\\
Z & 03:03:30.94 & $-$59:18:47.83 & 3.3 $\pm$ 1.0 & $<$4.7 & 46.6 $\pm$ 3.6 & 41.5 $\pm$ 3.2 & 24.4 $\pm$ 3.7\\

%% file: 0311/latexG
A$^{1}$ & 03:11:32.76 & $-$58:23:33.78 & 45.4 $\pm$ 1.6 & $45.0^{+1.6}_{-1.6}$ & 25.2 $\pm$ 4.0 & 38.1 $\pm$ 5.1 & 31.2 $\pm$ 6.1\\
B & 03:11:33.32 & $-$58:25:36.78 & 15.9 $\pm$ 1.5 & $15.2^{+1.6}_{-1.6}$ & 54.8 $\pm$ 7.7 & 35.9 $\pm$ 6.0 & 16.4 $\pm$ 7.1\\
C & 03:11:13.65 & $-$58:22:57.33 & 14.7 $\pm$ 1.8 & $13.8^{+1.8}_{-1.8}$ & 19.2 $\pm$ 7.7 & 16.7 $\pm$ 6.0 & 11.3 $\pm$ 7.5\\
D & 03:11:37.39 & $-$58:24:33.00 & 10.8 $\pm$ 1.5 & $10.0^{+1.5}_{-1.5}$ & 29.8 $\pm$ 8.0 & 32.8 $\pm$ 6.1 & 18.1 $\pm$ 7.2\\
E$^{1}$ & 03:11:30.45 & $-$58:23:20.11 & 9.6 $\pm$ 1.6 & $8.5^{+1.6}_{-1.7}$ & $--$ & $--$ & $--$\\
F$^{1}$ & 03:11:34.50 & $-$58:23:33.78 & 8.6 $\pm$ 1.6 & $7.4^{+1.7}_{-1.7}$ & $--$ & $--$ & $--$\\
G & 03:11:39.72 & $-$58:23:24.67 & 7.9 $\pm$ 1.6 & $6.7^{+1.7}_{-6.6}$ & 26.5 $\pm$ 7.2 & 19.5 $\pm$ 6.3 & 26.4 $\pm$ 7.2\\
\hline \multicolumn{8}{c}{Sources within 240\,arcsec and deboosted to zero}\\  \hline H & 03:11:10.70 & $-$58:25:14.00 & 9.2 $\pm$ 1.9 & $<$13.0 & 32.9 $\pm$ 8.8 & 22.5 $\pm$ 6.7 & 14.9 $\pm$ 8.0\\
I & 03:11:48.40 & $-$58:26:04.11 & 6.9 $\pm$ 1.6 & $<$9.9 & 9.9 $\pm$ 8.5 & 7.6 $\pm$ 7.6 & 3.5 $\pm$ 8.1\\
J$^{1}$ & 03:11:21.71 & $-$58:25:50.45 & 6.8 $\pm$ 1.7 & $<$9.8 & $--$ & $--$ & $--$\\
K & 03:11:32.18 & $-$58:24:14.78 & 6.6 $\pm$ 1.5 & $<$9.5 & 30.0 $\pm$ 7.9 & 38.4 $\pm$ 5.9 & 41.2 $\pm$ 7.3\\
\hline \multicolumn{8}{c}{Sources outside 240\,arcsec}\\  \hline L & 03:12:00.00 & $-$58:21:17.11 & 15.0 $\pm$ 2.3 & $13.5^{+2.3}_{-2.4}$ & 32.9 $\pm$ 8.4 & 28.1 $\pm$ 6.7 & 34.0 $\pm$ 9.2\\
M & 03:12:16.78 & $-$58:20:49.78 & 17.8 $\pm$ 4.7 & $<$20.3 & 17.8 $\pm$ 8.4 & 36.1 $\pm$ 8.0 & 23.5 $\pm$ 9.4\\
\hline \multicolumn{8}{c}{Sources detected at 3$\sigma$}\\  \hline N & 03:12:01.73 & $-$58:18:01.22 & 15.2 $\pm$ 4.9 & $<$13.4 & 0.2 $\pm$ 14.0 & $<$9.7 & $<$12.3\\
O & 03:11:50.13 & $-$58:28:29.89 & 10.9 $\pm$ 3.1 & $<$13.2 & 57.4 $\pm$ 11.9 & 58.0 $\pm$ 7.6 & 51.5 $\pm$ 9.8\\
P & 03:11:41.51 & $-$58:17:06.55 & 11.1 $\pm$ 3.4 & $<$12.1 & 19.8 $\pm$ 14.3 & 18.8 $\pm$ 20.8 & \dots\\
Q & 03:12:20.87 & $-$58:24:10.22 & 11.6 $\pm$ 3.7 & $<$12.1 & $<$11.4 & $<$8.8 & 3.7 $\pm$ 13.2\\
R & 03:11:39.68 & $-$58:27:21.56 & 6.1 $\pm$ 2.0 & $<$7.8 & 37.4 $\pm$ 10.1 & 28.9 $\pm$ 7.2 & 35.4 $\pm$ 8.7\\

%% file: 0348/latexG
A$^{1}$ & 03:48:41.89 & $-$62:20:52.05 & 39.9 $\pm$ 1.3 & $39.6^{+1.2}_{-1.2}$ & 53.2 $\pm$ 5.5 & 34.7 $\pm$ 5.3 & 28.8 $\pm$ 5.5\\
B & 03:49:00.87 & $-$62:21:10.28 & 11.0 $\pm$ 1.3 & $10.4^{+1.4}_{-1.4}$ & 19.0 $\pm$ 6.8 & 12.3 $\pm$ 4.8 & $<$6.2\\
C & 03:48:26.83 & $-$62:21:46.72 & 9.9 $\pm$ 1.2 & $9.4^{+1.2}_{-1.2}$ & 17.5 $\pm$ 6.0 & 21.3 $\pm$ 5.2 & 10.1 $\pm$ 5.8\\
D & 03:48:14.40 & $-$62:21:10.28 & 9.2 $\pm$ 1.5 & $8.2^{+1.6}_{-1.6}$ & 18.3 $\pm$ 6.1 & 18.2 $\pm$ 4.4 & 9.4 $\pm$ 6.3\\
E$^{1}$ & 03:48:43.86 & $-$62:20:20.17 & 8.5 $\pm$ 1.4 & $7.6^{+1.4}_{-1.5}$ & $--$ & $--$ & $--$\\
\hline \multicolumn{8}{c}{Sources within 240\,arcsec and deboosted to zero}\\  \hline F & 03:48:45.83 & $-$62:18:58.16 & 6.0 $\pm$ 1.5 & $<$8.6 & 37.4 $\pm$ 6.4 & 37.7 $\pm$ 4.4 & 42.4 $\pm$ 5.6\\
G & 03:48:43.20 & $-$62:20:42.94 & 5.9 $\pm$ 1.3 & $<$8.5 & 70.6 $\pm$ 6.4 & 55.3 $\pm$ 5.3 & 29.2 $\pm$ 5.9\\
H$^{1}$ & 03:48:32.04 & $-$62:23:58.83 & 5.5 $\pm$ 1.3 & $<$8.0 & $--$ & $--$ & $--$\\
\hline \multicolumn{8}{c}{Sources detected at 3$\sigma$}\\  \hline I & 03:49:47.99 & $-$62:20:47.50 & 16.4 $\pm$ 4.6 & $<$18.0 & \dots & \dots & \dots\\
J & 03:49:13.25 & $-$62:15:24.05 & 8.6 $\pm$ 2.6 & $<$10.5 & 26.0 $\pm$ 10.0 & 13.5 $\pm$ 8.7 & 14.2 $\pm$ 9.3\\
K & 03:48:36.05 & $-$62:16:46.05 & 6.7 $\pm$ 2.1 & $<$8.4 & 33.0 $\pm$ 7.1 & 18.6 $\pm$ 5.7 & 15.3 $\pm$ 7.4\\
L & 03:49:11.33 & $-$62:19:30.05 & 5.7 $\pm$ 1.5 & $<$8.0 & 27.3 $\pm$ 6.7 & 39.6 $\pm$ 4.9 & 36.6 $\pm$ 6.1\\
M & 03:49:27.71 & $-$62:21:28.50 & 5.6 $\pm$ 1.8 & $<$7.3 & $<$8.7 & $<$9.7 & $<$10.9\\
N & 03:49:08.08 & $-$62:21:55.83 & 5.0 $\pm$ 1.4 & $<$7.0 & 7.0 $\pm$ 6.7 & 2.5 $\pm$ 5.5 & $<$7.1\\
O & 03:48:45.18 & $-$62:17:58.94 & 5.1 $\pm$ 1.6 & $<$6.8 & 13.6 $\pm$ 6.4 & 14.9 $\pm$ 5.3 & 5.3 $\pm$ 7.1\\
P & 03:48:28.77 & $-$62:23:36.06 & 4.4 $\pm$ 1.3 & $<$6.2 & $<$6.6 & 2.6 $\pm$ 4.7 & 1.8 $\pm$ 6.2\\

%% file: 0457/latexG
A$^{1}$ & 04:57:17.21 & $-$49:31:55.06 & 33.9 $\pm$ 1.4 & $33.6^{+1.4}_{-1.4}$ & 78.4 $\pm$ 4.5 & 62.1 $\pm$ 2.9 & 49.4 $\pm$ 3.4\\
B & 04:57:24.23 & $-$49:32:26.94 & 14.8 $\pm$ 1.3 & $14.4^{+1.3}_{-1.3}$ & 16.6 $\pm$ 3.9 & 6.2 $\pm$ 2.7 & 8.1 $\pm$ 3.4\\
C & 04:57:08.34 & $-$49:29:15.61 & 10.6 $\pm$ 1.8 & $9.4^{+1.8}_{-1.9}$ & 57.5 $\pm$ 4.3 & 54.7 $\pm$ 3.1 & 41.9 $\pm$ 3.7\\
D$^{1}$ & 04:57:19.09 & $-$49:32:04.17 & 9.2 $\pm$ 1.3 & $8.4^{+1.4}_{-1.4}$ & $--$ & $--$ & $--$\\
\hline \multicolumn{8}{c}{Sources within 240\,arcsec and deboosted to zero}\\  \hline E & 04:57:27.98 & $-$49:31:23.17 & 6.5 $\pm$ 1.4 & $<$9.2 & 40.9 $\pm$ 3.9 & 41.2 $\pm$ 2.9 & 32.3 $\pm$ 3.6\\
\hline \multicolumn{8}{c}{Sources outside 240\,arcsec}\\  \hline F & 04:57:01.80 & $-$49:28:11.83 & 13.9 $\pm$ 2.4 & $12.0^{+2.6}_{-2.6}$ & 60.6 $\pm$ 4.4 & 66.5 $\pm$ 3.4 & 48.0 $\pm$ 4.5\\
G & 04:58:06.36 & $-$49:32:40.61 & 22.9 $\pm$ 4.3 & $<$30.8 & \dots & \dots & \dots\\
H & 04:56:53.41 & $-$49:26:27.05 & 18.1 $\pm$ 4.8 & $<$20.4 & 19.2 $\pm$ 4.5 & 6.9 $\pm$ 3.3 & $<$4.9\\
I & 04:57:47.60 & $-$49:28:30.05 & 12.3 $\pm$ 2.5 & $<$17.2 & 44.8 $\pm$ 4.6 & 34.6 $\pm$ 3.3 & 31.0 $\pm$ 4.5\\
J & 04:57:53.70 & $-$49:30:42.17 & 10.9 $\pm$ 2.6 & $<$15.0 & 20.6 $\pm$ 4.8 & 16.4 $\pm$ 5.7 & 14.1 $\pm$ 6.6\\
\hline \multicolumn{8}{c}{Sources detected at 3$\sigma$}\\  \hline K & 04:57:22.37 & $-$49:24:24.05 & 12.3 $\pm$ 3.8 & $<$12.8 & \dots & 14.7 $\pm$ 9.9 & 2.0 $\pm$ 8.0\\
L & 04:57:54.68 & $-$49:35:06.39 & 9.6 $\pm$ 3.0 & $<$10.9 & 5.9 $\pm$ 5.3 & 20.5 $\pm$ 4.3 & 27.5 $\pm$ 6.6\\
M & 04:57:05.50 & $-$49:34:39.06 & 5.7 $\pm$ 1.7 & $<$7.8 & 40.1 $\pm$ 3.9 & 53.4 $\pm$ 2.9 & 41.1 $\pm$ 3.6\\
N & 04:57:08.78 & $-$49:33:58.06 & 5.1 $\pm$ 1.5 & $<$7.2 & 3.5 $\pm$ 3.7 & $<$2.9 & 2.2 $\pm$ 3.5\\
O & 04:57:08.77 & $-$49:35:20.06 & 5.5 $\pm$ 1.8 & $<$7.1 & $<$5.1 & 7.1 $\pm$ 3.1 & 9.0 $\pm$ 3.9\\
P & 04:57:17.68 & $-$49:31:41.39 & 4.8 $\pm$ 1.4 & $<$6.7 & 67.2 $\pm$ 3.8 & 62.4 $\pm$ 3.2 & 49.0 $\pm$ 3.4\\
Q & 04:57:16.27 & $-$49:35:20.06 & 4.9 $\pm$ 1.6 & $<$6.6 & 1.9 $\pm$ 3.6 & 11.4 $\pm$ 2.8 & 10.2 $\pm$ 3.7\\
R & 04:57:10.19 & $-$49:32:13.28 & 4.5 $\pm$ 1.4 & $<$6.1 & 38.5 $\pm$ 3.5 & 25.0 $\pm$ 3.0 & 9.1 $\pm$ 3.9\\

%% file: 0553/latexG
A$^{1}$ & 05:53:20.21 & $-$50:07:10.61 & 45.2 $\pm$ 1.5 & $44.9^{+1.5}_{-1.5}$ & 57.7 $\pm$ 5.5 & 33.0 $\pm$ 5.1 & 32.3 $\pm$ 5.5\\
B & 05:53:21.63 & $-$50:09:04.50 & 19.1 $\pm$ 1.4 & $18.6^{+1.4}_{-1.4}$ & 56.2 $\pm$ 6.2 & 49.6 $\pm$ 4.9 & 36.2 $\pm$ 5.3\\
C & 05:53:06.95 & $-$50:07:33.39 & 12.7 $\pm$ 1.4 & $12.1^{+1.4}_{-1.4}$ & 48.7 $\pm$ 6.6 & 53.2 $\pm$ 5.3 & 32.4 $\pm$ 6.0\\
D$^{1}$ & 05:53:20.68 & $-$50:07:28.83 & 9.9 $\pm$ 1.5 & $9.1^{+1.5}_{-1.5}$ & $--$ & $--$ & $--$\\
E & 05:53:11.69 & $-$50:05:39.50 & 9.1 $\pm$ 1.5 & $8.2^{+1.5}_{-1.6}$ & 11.3 $\pm$ 5.9 & 23.5 $\pm$ 4.6 & 18.6 $\pm$ 5.6\\
F$^{1}$ & 05:53:16.89 & $-$50:07:37.95 & 7.6 $\pm$ 1.4 & $6.6^{+1.5}_{-1.5}$ & $--$ & $--$ & $--$\\
\hline \multicolumn{8}{c}{Sources within 240\,arcsec and deboosted to zero}\\  \hline G & 05:53:01.73 & $-$50:08:28.06 & 7.7 $\pm$ 1.7 & $<$11.0 & 26.5 $\pm$ 7.0 & 30.5 $\pm$ 4.9 & 13.8 $\pm$ 7.0\\
\hline \multicolumn{8}{c}{Sources outside 240\,arcsec}\\  \hline H & 05:53:17.83 & $-$50:13:46.95 & 23.3 $\pm$ 5.5 & $<$27.7 & 35.4 $\pm$ 8.5 & 36.2 $\pm$ 9.5 & 24.3 $\pm$ 8.6\\
\hline \multicolumn{8}{c}{Sources detected at 3$\sigma$}\\  \hline I & 05:53:56.26 & $-$50:12:06.73 & 11.5 $\pm$ 3.8 & $<$11.4 & $<$7.6 & 0.2 $\pm$ 6.4 & $<$7.1\\
J & 05:53:46.28 & $-$50:08:50.83 & 7.7 $\pm$ 2.2 & $<$10.1 & $<$6.6 & $<$6.1 & $<$7.8\\
K & 05:53:09.82 & $-$50:01:24.39 & 9.1 $\pm$ 3.0 & $<$10.0 & 11.8 $\pm$ 9.3 & 2.0 $\pm$ 7.7 & 1.5 $\pm$ 11.3\\
L & 05:53:19.75 & $-$50:01:10.72 & 8.5 $\pm$ 2.8 & $<$9.6 & 11.1 $\pm$ 11.9 & \dots & $<$11.7\\
M & 05:53:13.59 & $-$50:04:49.39 & 5.4 $\pm$ 1.6 & $<$7.4 & 27.5 $\pm$ 6.2 & 30.7 $\pm$ 4.7 & 31.1 $\pm$ 6.4\\
N & 05:53:28.74 & $-$50:06:20.50 & 5.2 $\pm$ 1.7 & $<$6.7 & 8.1 $\pm$ 5.9 & 19.5 $\pm$ 4.6 & 21.0 $\pm$ 5.9\\
O & 05:53:04.11 & $-$50:06:25.06 & 4.7 $\pm$ 1.5 & $<$6.4 & 40.0 $\pm$ 6.2 & 25.9 $\pm$ 4.9 & 27.7 $\pm$ 5.4\\

%% file: 2018/latexG
A$^{1}$ & 20:19:02.99 & $-$45:05:08.72 & 26.2 $\pm$ 1.4 & $25.8^{+1.4}_{-1.4}$ & 89.7 $\pm$ 5.4 & 86.5 $\pm$ 3.8 & 59.7 $\pm$ 4.9\\
B$^{1}$ & 20:18:56.11 & $-$45:04:50.50 & 13.7 $\pm$ 1.5 & $13.0^{+1.5}_{-1.5}$ & $--$ & $--$ & $--$\\
C$^{1}$ & 20:19:00.41 & $-$45:04:27.72 & 9.0 $\pm$ 1.4 & $8.2^{+1.5}_{-1.5}$ & $--$ & $--$ & $--$\\
D$^{1}$ & 20:18:57.83 & $-$45:04:32.28 & 8.7 $\pm$ 1.4 & $7.7^{+1.5}_{-1.5}$ & $--$ & $--$ & $--$\\
E & 20:19:10.73 & $-$45:03:23.94 & 8.2 $\pm$ 1.6 & $7.1^{+1.7}_{-1.7}$ & 18.1 $\pm$ 5.5 & 19.0 $\pm$ 4.9 & 13.5 $\pm$ 6.1\\
F$^{1}$ & 20:19:02.56 & $-$45:04:45.94 & 7.7 $\pm$ 1.4 & $6.7^{+1.5}_{-1.5}$ & $--$ & $--$ & $--$\\
\hline \multicolumn{8}{c}{Sources within 240\,arcsec and deboosted to zero}\\  \hline G$^{1}$ & 20:19:04.71 & $-$45:05:17.83 & 6.6 $\pm$ 1.4 & $<$9.4 & $--$ & $--$ & $--$\\
\hline \multicolumn{8}{c}{Sources outside 240\,arcsec}\\  \hline H & 20:19:19.75 & $-$45:02:11.05 & 10.2 $\pm$ 1.9 & $8.7^{+2.0}_{-2.1}$ & $<$7.9 & 10.4 $\pm$ 5.6 & 7.5 $\pm$ 8.3\\
I & 20:18:24.30 & $-$45:03:19.39 & 14.4 $\pm$ 3.8 & $<$17.7 & \dots & 30.9 $\pm$ 8.5 & 12.8 $\pm$ 14.1\\
J & 20:18:47.49 & $-$45:08:24.61 & 11.6 $\pm$ 2.9 & $<$15.4 & 20.8 $\pm$ 7.6 & 20.5 $\pm$ 5.4 & 10.7 $\pm$ 7.9\\
\hline \multicolumn{8}{c}{Sources detected at 3$\sigma$}\\  \hline K & 20:19:22.80 & $-$45:10:13.95 & 16.2 $\pm$ 5.2 & $<$13.7 & $<$7.9 & $<$4.9 & $<$8.1\\
L & 20:18:37.60 & $-$45:06:07.95 & 9.7 $\pm$ 2.7 & $<$12.5 & $<$6.0 & $<$4.9 & $<$6.7\\
M & 20:19:03.42 & $-$45:00:03.50 & 6.5 $\pm$ 1.8 & $<$8.8 & $<$8.6 & 18.9 $\pm$ 6.6 & 5.8 $\pm$ 8.1\\
N & 20:19:06.86 & $-$45:06:03.39 & 5.4 $\pm$ 1.5 & $<$7.6 & 22.1 $\pm$ 6.2 & 16.1 $\pm$ 4.4 & 12.1 $\pm$ 5.6\\
O & 20:18:56.97 & $-$45:03:42.17 & 4.9 $\pm$ 1.5 & $<$6.9 & 34.7 $\pm$ 6.5 & 55.2 $\pm$ 4.7 & 52.4 $\pm$ 5.2\\
P & 20:18:54.83 & $-$45:01:48.28 & 5.0 $\pm$ 1.7 & $<$6.6 & $<$5.6 & 4.3 $\pm$ 4.8 & 2.4 $\pm$ 5.5\\
Q & 20:18:54.82 & $-$45:04:59.61 & 4.7 $\pm$ 1.5 & $<$6.4 & 91.5 $\pm$ 6.2 & 100.2 $\pm$ 4.8 & 125.1 $\pm$ 6.0\\
R & 20:18:58.69 & $-$45:05:04.17 & 4.4 $\pm$ 1.4 & $<$5.9 & 46.0 $\pm$ 6.1 & 42.0 $\pm$ 5.0 & 11.5 $\pm$ 6.1\\

%% file: 2052/latexG
A$^{1}$ & 20:52:41.44 & $-$56:11:49.72 & 20.0 $\pm$ 1.2 & $19.7^{+1.2}_{-1.2}$ & 63.7 $\pm$ 6.0 & 49.3 $\pm$ 5.2 & 13.4 $\pm$ 5.4\\
B & 20:52:29.99 & $-$56:11:04.17 & 13.0 $\pm$ 1.3 & $12.5^{+1.3}_{-1.3}$ & 30.2 $\pm$ 5.7 & 35.2 $\pm$ 4.5 & 23.2 $\pm$ 5.4\\
C & 20:52:43.07 & $-$56:14:56.50 & 11.3 $\pm$ 1.2 & $10.8^{+1.2}_{-1.2}$ & 52.9 $\pm$ 6.7 & 66.8 $\pm$ 4.7 & 75.6 $\pm$ 7.1\\
D$^{1}$ & 20:52:41.99 & $-$56:12:07.95 & 8.8 $\pm$ 1.1 & $8.2^{+1.2}_{-1.2}$ & $--$ & $--$ & $--$\\
E & 20:52:19.58 & $-$56:13:43.61 & 8.7 $\pm$ 1.5 & $7.7^{+1.5}_{-1.6}$ & 35.9 $\pm$ 6.8 & 38.6 $\pm$ 5.1 & 12.7 $\pm$ 5.8\\
F & 20:52:28.33 & $-$56:13:02.61 & 6.4 $\pm$ 1.2 & $5.6^{+1.3}_{-1.3}$ & 16.8 $\pm$ 5.0 & 38.5 $\pm$ 4.4 & 30.7 $\pm$ 6.3\\
\hline \multicolumn{8}{c}{Sources within 240\,arcsec and deboosted to zero}\\  \hline G$^{1}$ & 20:52:40.90 & $-$56:11:31.50 & 5.3 $\pm$ 1.2 & $<$7.6 & $--$ & $--$ & $--$\\
\hline \multicolumn{8}{c}{Sources outside 240\,arcsec}\\  \hline H & 20:52:48.01 & $-$56:05:58.94 & 16.9 $\pm$ 1.8 & $16.1^{+1.8}_{-1.8}$ & 44.8 $\pm$ 8.7 & 50.6 $\pm$ 9.1 & 33.9 $\pm$ 8.5\\
I & 20:52:32.75 & $-$56:07:11.83 & 9.7 $\pm$ 1.8 & $8.4^{+1.9}_{-1.9}$ & 18.2 $\pm$ 8.6 & 25.1 $\pm$ 8.4 & 13.6 $\pm$ 7.7\\
\hline \multicolumn{8}{c}{Sources detected at 3$\sigma$}\\  \hline J & 20:52:03.19 & $-$56:13:48.17 & 11.8 $\pm$ 3.5 & $<$13.3 & 8.1 $\pm$ 7.1 & $<$7.4 & 2.3 $\pm$ 7.3\\
K & 20:52:38.22 & $-$56:03:51.39 & 14.2 $\pm$ 4.7 & $<$12.1 & \dots & $<$17.3 & \dots\\
L & 20:52:16.37 & $-$56:09:14.83 & 6.5 $\pm$ 1.9 & $<$8.6 & 1.8 $\pm$ 6.7 & 19.5 $\pm$ 5.8 & 16.5 $\pm$ 7.0\\
M & 20:53:02.17 & $-$56:07:25.50 & 5.5 $\pm$ 1.6 & $<$7.5 & 25.0 $\pm$ 8.1 & 29.4 $\pm$ 5.6 & 15.7 $\pm$ 6.3\\
N & 20:52:26.76 & $-$56:07:20.94 & 5.7 $\pm$ 1.9 & $<$7.3 & 3.8 $\pm$ 8.4 & $<$7.1 & 0.9 $\pm$ 7.5\\
O & 20:52:21.76 & $-$56:14:10.95 & 5.0 $\pm$ 1.5 & $<$7.0 & 37.4 $\pm$ 6.3 & 44.3 $\pm$ 5.1 & 39.5 $\pm$ 6.1\\
P & 20:52:49.64 & $-$56:08:11.05 & 4.6 $\pm$ 1.4 & $<$6.2 & 21.5 $\pm$ 6.8 & 20.1 $\pm$ 5.8 & 22.1 $\pm$ 6.8\\
Q & 20:52:57.82 & $-$56:08:01.94 & 4.6 $\pm$ 1.5 & $<$6.1 & $<$6.7 & $<$5.9 & $<$6.4\\

%% file: 2335/latexG
A$^{1}$ & 23:35:13.30 & $-$53:24:27.33 & 29.5 $\pm$ 1.4 & $29.1^{+1.4}_{-1.4}$ & 63.4 $\pm$ 7.2 & 48.4 $\pm$ 6.6 & 27.5 $\pm$ 6.7\\
B & 23:35:21.46 & $-$53:21:34.22 & 12.1 $\pm$ 1.7 & $11.0^{+1.8}_{-1.8}$ & 12.8 $\pm$ 10.0 & 16.2 $\pm$ 8.5 & 6.9 $\pm$ 8.9\\
C & 23:35:28.59 & $-$53:24:22.78 & 10.4 $\pm$ 1.5 & $9.6^{+1.6}_{-1.6}$ & 18.3 $\pm$ 9.1 & 29.7 $\pm$ 8.1 & 16.0 $\pm$ 7.9\\
D & 23:35:19.41 & $-$53:24:36.44 & 8.7 $\pm$ 1.4 & $7.7^{+1.5}_{-1.5}$ & 26.9 $\pm$ 10.2 & 30.4 $\pm$ 8.1 & 16.0 $\pm$ 8.7\\
E & 23:35:11.26 & $-$53:25:58.45 & 7.7 $\pm$ 1.3 & $6.9^{+1.4}_{-1.4}$ & 47.6 $\pm$ 9.8 & 68.8 $\pm$ 8.3 & 68.5 $\pm$ 8.8\\
\hline \multicolumn{8}{c}{Sources within 240\,arcsec and deboosted to zero}\\  \hline F & 23:35:19.41 & $-$53:25:08.33 & 6.0 $\pm$ 1.4 & $<$8.7 & 16.0 $\pm$ 9.3 & $<$8.2 & $<$8.8\\
\hline \multicolumn{8}{c}{Sources outside 240\,arcsec}\\  \hline G & 23:34:38.15 & $-$53:24:09.11 & 17.8 $\pm$ 4.2 & $<$22.6 & $<$9.7 & 2.0 $\pm$ 7.8 & 0.8 $\pm$ 8.9\\
H & 23:35:25.02 & $-$53:18:18.33 & 15.3 $\pm$ 3.3 & $<$21.0 & \dots & \dots & \dots\\
I & 23:35:38.29 & $-$53:29:05.23 & 10.0 $\pm$ 2.4 & $<$14.0 & 42.2 $\pm$ 12.1 & 26.2 $\pm$ 9.2 & 18.9 $\pm$ 9.4\\
\hline \multicolumn{8}{c}{Sources detected at 3$\sigma$}\\  \hline J & 23:35:46.37 & $-$53:19:40.33 & 17.8 $\pm$ 5.0 & $<$18.3 & 9.1 $\pm$ 10.6 & 27.3 $\pm$ 9.6 & 13.8 $\pm$ 9.6\\
K & 23:34:43.32 & $-$53:19:49.44 & 17.6 $\pm$ 5.3 & $<$15.9 & 76.4 $\pm$ 12.0 & 25.7 $\pm$ 9.5 & 14.3 $\pm$ 9.5\\
L & 23:35:36.22 & $-$53:22:24.33 & 7.1 $\pm$ 2.0 & $<$9.5 & 16.9 $\pm$ 9.7 & 1.5 $\pm$ 8.2 & $<$8.1\\
M & 23:35:06.20 & $-$53:19:54.00 & 6.8 $\pm$ 2.2 & $<$8.3 & 28.9 $\pm$ 11.3 & 19.2 $\pm$ 9.5 & 1.6 $\pm$ 9.4\\
N & 23:35:14.84 & $-$53:20:12.22 & 6.3 $\pm$ 2.0 & $<$8.1 & 16.8 $\pm$ 11.7 & 21.2 $\pm$ 9.3 & 14.9 $\pm$ 9.8\\
O & 23:35:06.20 & $-$53:20:30.44 & 6.2 $\pm$ 2.0 & $<$8.0 & 10.6 $\pm$ 10.3 & 10.2 $\pm$ 9.2 & 3.2 $\pm$ 9.0\\
P & 23:35:14.82 & $-$53:28:10.56 & 5.6 $\pm$ 1.7 & $<$7.6 & 20.3 $\pm$ 10.7 & 32.6 $\pm$ 9.0 & 21.7 $\pm$ 9.7\\
Q & 23:35:13.30 & $-$53:24:00.00 & 5.2 $\pm$ 1.5 & $<$7.3 & 86.9 $\pm$ 9.8 & 54.7 $\pm$ 8.1 & 28.6 $\pm$ 9.0\\

%% file: 2349/latexG
A$^{1}$ & 23:49:42.68 & $-$56:38:23.50 & 59.2 $\pm$ 1.4 & $58.9^{+1.4}_{-1.4}$ & 96.0 $\pm$ 5.7 & 80.8 $\pm$ 5.0 & 40.8 $\pm$ 5.5\\
B$^2$ & 23:49:43.23 & $-$56:37:28.83 & 21.9 $\pm$ 1.4 & $21.5^{+1.4}_{-1.4}$ & 15.1 $\pm$ 5.5 & 16.6 $\pm$ 5.3 & 20.0 $\pm$ 5.7\\
C$^{1}$ & 23:49:41.02 & $-$56:38:05.28 & 11.2 $\pm$ 1.4 & $10.4^{+1.5}_{-1.5}$ & $--$ & $--$ & $--$\\
D$^{1}$ & 23:49:44.89 & $-$56:38:18.95 & 9.0 $\pm$ 1.4 & $8.2^{+1.5}_{-1.5}$ & $--$ & $--$ & $--$\\
E & 23:50:08.63 & $-$56:38:14.39 & 8.9 $\pm$ 1.7 & $7.5^{+1.9}_{-7.4}$ & 15.2 $\pm$ 7.7 & 22.9 $\pm$ 5.6 & 20.5 $\pm$ 7.6\\
F$^{1}$ & 23:49:43.78 & $-$56:38:46.28 & 7.8 $\pm$ 1.4 & $6.7^{+1.5}_{-1.6}$ & $--$ & $--$ & $--$\\
\hline \multicolumn{8}{c}{Sources within 240\,arcsec and deboosted to zero}\\  \hline G & 23:49:41.58 & $-$56:37:42.50 & 6.0 $\pm$ 1.4 & $<$8.6 & 58.3 $\pm$ 5.8 & 32.8 $\pm$ 5.2 & 14.5 $\pm$ 5.9\\
H$^2$ & 23:49:33.85 & $-$56:37:47.06 & 5.3 $\pm$ 1.4 & $<$7.5 & $--$ & $--$ & $--$\\
\hline \multicolumn{8}{c}{Sources detected at 3$\sigma$}\\  \hline I & 23:50:42.84 & $-$56:36:34.17 & 26.9 $\pm$ 7.6 & $<$20.2 & 7.1 $\pm$ 12.8 & 16.4 $\pm$ 10.1 & 11.6 $\pm$ 10.7\\
J & 23:49:38.21 & $-$56:45:22.62 & 20.0 $\pm$ 6.0 & $<$16.2 & \dots & \dots & 2.1 $\pm$ 10.5\\
K & 23:48:49.69 & $-$56:37:15.17 & 21.8 $\pm$ 6.7 & $<$15.6 & \dots & \dots & \dots\\
L & 23:50:36.75 & $-$56:35:48.61 & 17.8 $\pm$ 5.5 & $<$14.7 & \dots & \dots & \dots\\
M & 23:50:26.34 & $-$56:40:40.17 & 10.6 $\pm$ 3.3 & $<$12.0 & $<$13.5 & 4.8 $\pm$ 9.3 & $<$11.9\\
N & 23:49:37.67 & $-$56:43:33.28 & 8.6 $\pm$ 2.8 & $<$9.8 & $<$8.9 & $<$8.0 & $<$10.4\\
O & 23:49:13.37 & $-$56:40:12.84 & 7.6 $\pm$ 2.4 & $<$9.3 & 6.7 $\pm$ 7.5 & $<$5.2 & $<$8.5\\
P & 23:49:42.16 & $-$56:32:09.94 & 7.0 $\pm$ 2.1 & $<$9.2 & 32.6 $\pm$ 10.5 & 23.5 $\pm$ 9.3 & 1.1 $\pm$ 14.7\\
Q & 23:49:50.97 & $-$56:34:08.39 & 6.0 $\pm$ 1.7 & $<$8.3 & 16.6 $\pm$ 8.5 & 10.2 $\pm$ 6.0 & 13.6 $\pm$ 7.4\\
R & 23:50:04.76 & $-$56:35:53.17 & 6.1 $\pm$ 1.8 & $<$8.1 & $<$6.6 & $<$4.9 & $<$5.9\\
S & 23:49:48.19 & $-$56:42:29.50 & 6.0 $\pm$ 2.0 & $<$7.5 & $<$10.0 & 3.0 $\pm$ 6.3 & $<$6.8\\
T & 23:49:40.47 & $-$56:38:28.06 & 4.7 $\pm$ 1.4 & $<$6.6 & 91.7 $\pm$ 6.4 & 40.2 $\pm$ 5.3 & 9.2 $\pm$ 5.8\\